\documentclass[longauth,traditabstract,usenatbib]{aa}
\def\del#1{}  \def\remark#1{}

\def\symbolfootnote[#1]#2{\begingroup\def\thefootnote{\fnsymbol{footnote}}
\footnote[#1]{#2}\endgroup}
\newcommand{\rmn}{\mathrm}
\newcommand{\CR}{\mathrm{CR}}

\newcommand{\dps}{\displaystyle}
\newcommand{\eps}{\varepsilon}
\usepackage{graphicx}
\usepackage{epstopdf}
\usepackage{natbib}
\usepackage{txfonts}
\usepackage{color}

\begin{document}

%%%%%%%%%%%%%%%%%%%%%%%%%%%%%%%%%%%%%%%%
\title{Deep observation of the NGC\,1275 region with MAGIC:
search of diffuse $\gamma$-ray emission from cosmic rays in the Perseus cluster}

%%%%%%%%%%%%%%%%%%%%%%%%%%%%%%%%%%%%%%%%
\author{
M.~L.~Ahnen\inst{1} \and
S.~Ansoldi\inst{2} \and
L.~A.~Antonelli\inst{3} \and
P.~Antoranz\inst{4} \and
A.~Babic\inst{5} \and
B.~Banerjee\inst{6} \and
P.~Bangale\inst{7} \and
U.~Barres de Almeida\inst{7,}\inst{25} \and
J.~A.~Barrio\inst{8} \and
J.~Becerra Gonz\'alez\inst{9,}\inst{26}
% ...
}
\institute {  ETH Zurich, CH-8093 Zurich, Switzerland
\and Universit\`a di Udine, and INFN Trieste, I-33100 Udine, Italy
\and INAF National Institute for Astrophysics, I-00136 Rome, Italy
\and Universit\`a  di Siena, and INFN Pisa, I-53100 Siena, Italy
\and Croatian MAGIC Consortium, Rudjer Boskovic Institute, University of Rijeka, University of Split and University of Zagreb, Croatia
\and Saha Institute of Nuclear Physics, 1/AF Bidhannagar, Salt Lake, Sector-1, Kolkata 700064, India
\and Max-Planck-Institut f\"ur Physik, D-80805 M\"unchen, Germany
\and Universidad Complutense, E-28040 Madrid, Spain
\and Inst. de Astrof\'isica de Canarias, E-38200 La Laguna, Tenerife, Spain; Universidad de La Laguna, Dpto. Astrof\'isica, E-38206 La Laguna, Tenerife, Spain
\and University of \L\'od\'z, PL-90236 Lodz, Poland
\and Deutsches Elektronen-Synchrotron (DESY), D-15738 Zeuthen, Germany
\and Institut de Fisica d'Altes Energies (IFAE), The Barcelona Institute of Science and Technology, Campus UAB, 08193 Bellaterra (Barcelona), Spain
\and Universit\"at W\"urzburg, D-97074 W\"urzburg, Germany
\and Universit\`a di Padova and INFN, I-35131 Padova, Italy
\and Centro de Investigaciones Energ\'eticas, Medioambientales y Tecnol\'ogicas, E-28040 Madrid, Spain
\and Institute for Space Sciences (CSIC/IEEC), E-08193 Barcelona, Spain
\and Technische Universit\"at Dortmund, D-44221 Dortmund, Germany
\and Unitat de F\'isica de les Radiacions, Departament de F\'isica, and CERES-IEEC, Universitat Aut\`onoma de Barcelona, E-08193 Bellaterra, Spain
\and Universitat de Barcelona, ICC, IEEC-UB, E-08028 Barcelona, Spain
\and Japanese MAGIC Consortium, ICRR, The University of Tokyo, Department of Physics and Hakubi Center, Kyoto University, Tokai University, The University of Tokushima, KEK, Japan
\and Finnish MAGIC Consortium, Tuorla Observatory, University of Turku and Department of Physics, University of Oulu, Finland
\and Inst. for Nucl. Research and Nucl. Energy, BG-1784 Sofia, Bulgaria
\and Universit\`a di Pisa, and INFN Pisa, I-56126 Pisa, Italy
\and ICREA and Institute for Space Sciences (CSIC/IEEC), E-08193 Barcelona, Spain
\and now at Centro Brasileiro de Pesquisas F\'isicas (CBPF/MCTI), R. Dr. Xavier Sigaud, 150 - Urca, Rio de Janeiro - RJ, 22290-180, Brazil
\and now at NASA Goddard Space Flight Center, Greenbelt, MD 20771, USA and Department of Physics and Department of Astronomy, University of Maryland, College Park, MD 20742, USA
% ...
}

%\date{Accepted XXX. Received XXX; in original from XXX}

%%%%%%%%%%%%%%%%%%%%%%%%%%%%%%%%%%%%%%%
\abstract{
Clusters of galaxies are expected to be reservoirs of cosmic rays (CRs) that should produce
diffuse $\gamma$-ray emission due to their hadronic interactions with the intra-cluster medium.
The nearby Perseus cool-core cluster, identified as the most promising target to search for such an 
emission, has been observed with the MAGIC telescopes
at very-high energies (VHE, $E\gtrsim100$\,GeV) for a total of 253\,hr from 2009 to 2014.
The active nuclei of NGC\,1275, the central dominant galaxy of the cluster, and IC\,310, lying at about $0.6^\circ$ from
the centre, have been detected as point-like VHE $\gamma$-ray emitters during the first phase of this campaign.
We report an updated measurement of the NGC\,1275 spectrum, which is described well
by a power law
with a photon index $\Gamma=3.6\pm0.2_{\rmn{stat}}\pm0.2_{\rmn{syst}}$ between 90\,GeV and 1200\,GeV.
We do not detect any diffuse $\gamma$-ray emission from the cluster and so set stringent constraints on its CR population.
To bracket the uncertainties over the CR spatial and spectral distributions, we adopt different spatial templates
and power-law spectral indexes $\alpha$.
For $\alpha=2.2$, the CR-to-thermal pressure within the cluster virial radius is constrained to be $\lesssim 1-2$\%,
except if CRs can propagate out of the cluster core, generating a flatter radial distribution 
and releasing the CR-to-thermal pressure constraint to $\lesssim 20$\%.
Assuming that the observed radio mini-halo of Perseus is generated by secondary electrons from CR hadronic interactions, we can derive lower limits
on the central magnetic field, $B_0$, that depend on the CR distribution. For $\alpha = 2.2$, $B_0 \gtrsim 5 - 8$\,$\mu$G,
which is below the $\sim$25\,$\mu$G inferred from Faraday rotation measurements, whereas
for $\alpha \lesssim 2.1$, the hadronic interpretation of the diffuse radio emission contrasts with
our $\gamma$-ray flux upper limits independently of the magnetic field strength.
}


\titlerunning{Search of diffuse $\gamma$-ray emissions from cosmic rays in the Perseus cluster}

\authorrunning{M.~L.~Ahnen et al.}

\maketitle

%%%%%%%%%%%%%%%%%%%%%%%%%%%%%%%%%%%%%%%%%%%%%%%%%%%%%%%%%%%%%%%%%%%
%%%%%%%%%%%%%%%%%%%%%%%%%%%%%%%%%%%%%%%%%%%%%%%%%%%%%%%%%%%%%%%%%%%
\section{Introduction}
\label{sec:1}
\begingroup
\let\thefootnote\relax\footnotetext{* Corresponding authors: P.~Colin 
(colin@mppmu.mpg.de) and F.~Zandanel (f.zandanel@uva.nl).\\}
\endgroup
Clusters of galaxies represent the latest stage of structure formation, 
and are presently assembling through mergers of smaller groups of 
galaxies and gas accretion. They are powerful cosmological tools 
for testing the evolution of the Universe (see \citealp{2005RvMP...77..207V} for a review). 
Cosmic-ray (CR) protons can accumulate in clusters of galaxies for cosmological times, 
accelerated by structure formation shocks, 
and outflows from active galactic nuclei (AGNs) and galaxies (see, e.g., 
\citealp{1996SSRv...75..279V,1997ApJ...487..529B}; 
see \citealp{2014IJMPD..2330007B} for a review). These CR protons can 
interact hadronically with the protons of the intra-cluster 
medium (ICM), a hot thermal plasma ($k_\rmn{B}T\simeq$\,1--10\,keV) filling the cluster volume, and
generate pions. While the charged pions decay 
to secondary electrons and neutrinos, the neutral pions decay directly 
to high-energy $\gamma$ rays.
Despite many observational efforts in the past decade, $\gamma$-ray 
emission from clusters of galaxies remains elusive.\footnote{For space-based cluster observations in the GeV-band, see 
\cite{2003ApJ...588..155R, 2010JCAP...05..025A,
2010ApJ...717L..71A,2011ApJ...728...53J,2012arXiv1207.6749H,2012JCAP...07..017A,
2013arXiv1308.6278H,2014MNRAS.440..663Z,
2013arXiv1308.5654T,2013arXiv1309.0197P,2014MNRAS.437.2291V,2014arXiv1405.7047G,
2014arXiv1410.4562S,2015arXiv150502782V,2015arXiv150708995T,2015arXiv151000004A}. 
For ground-based observations in the energy band 
above 100\,GeV, see \cite{2006ApJ...644..148P, 
2008AIPC.1085..569P,2009arXiv0907.0727T, 2009A&A...495...27A,
2009arXiv0907.3001D,2009arXiv0907.5000G,cangaroo_clusters, 
2009ApJ...706L.275A,2010ApJ...710..634A,2011arXiv1111.5544M,
2012...VERITAS,2012A&A...545A.103H}.}

Non-thermal emission is observed at radio frequencies in many clusters of 
galaxies in the form of diffuse synchrotron radiation
(see \citealp{2012A&ARv..20...54F} for a review). This probes for the presence 
of relativistic CR electrons and magnetic fields in the cluster environment.
However, a conclusive proof of CR-proton acceleration has yet to be found.
The observed CR electrons can also produce hard X-rays by inverse-Compton (IC) scattering of cosmic 
microwave background (CMB) photons. Several claims of IC detection have been made in the past (see 
\citealp{2008SSRv..134...71R} for a review), but more recent observations do not confirm them 
\citep{2009ApJ...690..367A,2010ApJ...725.1688A,2011ApJ...727..119W,
2012ApJ...748...67W,
2014ApJ...792...48W,2015ApJ...800..139G}, and the possible diffuse IC emission from 
clusters remains elusive, too.

The observed diffuse radio emission in clusters can be divided in two main 
categories: peripheral radio relics and central radio halos 
(e.g., \citealp{2012A&ARv..20...54F,2014IJMPD..2330007B}). The latter are 
usually divided in two other categories: giant-halos hosted in merging 
non-cool-core clusters (e.g., the Coma cluster; 
\citealp{1997A&A...321...55D,2011MNRAS.412....2B}), and mini-halos hosted in 
relaxed cool-core clusters (e.g., the Perseus cluster; 
\citealp{1990MNRAS.246..477P,Sijbring1993,2002A&A...386..456G}).
While radio relics can be roughly related to merger shocks due to their spatial coincidence and morphology,
the explanation for the origin of radio halos is more challenging.
The generation mechanism of radio halos has been historically debated 
between re-acceleration\footnote{See, e.g., 
\cite{1987A&A...182...21S,2001MNRAS.320..365B,2001ApJ...557..560P,
2002A&A...386..456G,2002ApJ...577..658O,2003ApJ...584..190F,2004MNRAS.350.1174B,
2005MNRAS.363.1173B,2005MNRAS.357.1313C,2007MNRAS.378..245B,2010arXiv1008.0184B,
2012arXiv1207.3025B,2013MNRAS.429.3564D,2013ApJ...762...78Z,2015arXiv150307870P,Miniati15a,2016MNRAS.455L..41B}
.} and hadronic models.\footnote{See, e.g., 
\cite{1980ApJ...239L..93D,1982AJ.....87.1266V,1997ApJ...477..560E,
1999APh....12..169B,2000A&A...362..151D, 2001ApJ...559...59M, 
2003MNRAS.342.1009M,2003A&A...407L..73P,2003APh....19..679G,2004A&A...413...17P, 
2004MNRAS.352...76P,2007IJMPA..22..681B, 2008MNRAS.385.1211P, 
2008MNRAS.385.1242P,2009JCAP...09..024K,2010MNRAS.401...47D, 
2010arXiv1003.0336D,2010arXiv1003.1133K,2010MNRAS.409..449P,2011arXiv1105.3240P,
2011A&A...527A..99E,2012ApJ...746...53F,2012arXiv1207.6410Z,2015ApJ...801..146Z}.} In the 
re-acceleration model, a seed population of CR electrons can be re-accelerated 
by interacting turbulent waves, while in the hadronic 
scenario, the radio-emitting electrons are secondaries produced by CR protons 
interacting with the protons of the ICM.
Currently, the re-acceleration scenario is favoured as the generation mechanism for 
giant radio halos, while for mini-halos, both the re-acceleration and hadronic 
models can explain the observed emission (see, e.g., 
\citealp{2011A&A...527A..99E} and \citealp{2014IJMPD..2330007B} for extensive 
discussions).

The key questions are the following. What is the origin of the 
radio-emitting electrons? What is the role of CR protons and how do they affect 
the cluster environment? Upcoming X-ray observations have the potential to 
detect IC emission in clusters and, hopefully, to break the degeneracy between 
the electron and the magnetic-field distributions 
\citep{2014arXiv1412.1176K,2015arXiv150106940B}, providing an alternative estimate 
for the magnetic field in clusters with respect to Faraday rotation (FR) measurements 
\citep{1991ApJ...379...80K,2001ApJ...547L.111C,2002ARA&A..40..319C,
2005A&A...434...67V,2010A&A...513A..30B,2011A&A...529A..13K,2013MNRAS.433.3208B}.
However, the presence and role of CR protons in clusters can
only be probed directly
through the $\gamma$ rays and neutrinos induced by hadronic interactions.  
The high-energy astronomy window is then crucial for understanding non-thermal phenomena in 
clusters of galaxies. (This is also true, albeit even more challenging, for neutrinos; see, e.g, 
\citealp{2008ApJ...689L.105M,2014arXiv1410.8697Z}.) 
 
The Perseus cluster of galaxies (a.k.a. Abell~426) is a relaxed cool-core
cluster located at a distance of about 
$D_\mathrm{lum}=78$\,Mpc (redshift $z=0.018$). It hosts the brightest thermal ICM in X-rays 
\citep{2002ApJ...567..716R} and a very luminous radio mini-halo 
\citep{1990MNRAS.246..477P,Sijbring1993,2002A&A...386..456G}. The high ICM 
density at the centre of the cluster \citep{2003ApJ...590..225C} implies a high 
density of target protons for hadronic interactions with CR protons. Therefore, 
Perseus is the best cluster for searching for CR-induced $\gamma$-ray emission 
(we refer the reader to 
\citealp{2010MNRAS.409..449P,2010ApJ...710..634A,2011arXiv1111.5544M,
2011arXiv1105.3240P} for a detailed argumentation).
The Perseus cluster also hosts three bright radio galaxies \citep{1968MNRS138}: NGC\,1275, the central dominant galaxy of the cluster,    
NGC\,1265, archetype of a head-tail radio galaxy, in which the jets are bent by their interaction with the ICM,
and IC\,310, a peculiar object that shows properties of different classifications and which could be an intermediate state
between a BL Lac and a radio galaxy. 
The AGNs of both NGC\,1275 and IC\,310 show a bright and variable 
$\gamma$-ray emission in the energy ranges of both the \emph{Fermi}-Large-Area-Telescope (LAT) \citep{2010A&A...519L...6N,2009arXiv0904.1904T}
and the atmospheric-Cherenkov telescopes \citep{2010ApJ...723L.207A,2012A&A...539L...2A, 2014A&A...564A...5A,2014A&A...563A..91A}.
The NGC\,1275 AGN emission obscures the expected diffuse cluster emission over most of the $\gamma$-ray band,
in particular in the GeV region, where the spectral energy distribution is peaking \citep{2014A&A...564A...5A}.
 
Since 2008, the Perseus cluster is intensively observed by the Major 
Atmospheric Gamma Imaging Cherenkov (MAGIC) telescopes. In this paper we present 
the results obtained with $\sim$250~hours of effective observation time taken in 
stereoscopic mode from 2009 to 2014 and derive tight constraints on the CR 
population in the cluster.
We discuss neither the AGN physics nor the possible indirect detection of dark matter 
from clusters \citep{2015ICRCPalacio}.
After describing the observation and data analysis in 
Section~\ref{sec:2}, we report our observational results and searches for 
a CR-induced signal in Sects.~\ref{sec:3} and~\ref{sec:4}, respectively. 
In Section~\ref{sec:5}, we discuss the interpretation of our constraints on the 
CR physics in galaxy clusters, and, finally, in Section~\ref{sec:6} we present 
our conclusions. Throughout the paper, we assume a standard $\Lambda$CDM 
cosmology with $H = 70$\,km\,s$^{-1}$\,Mpc$^{-1}$.

%%%%%%%%%%%%%%%%%%%%%%%%%%%%%%%%%%%%%%%%%%%%%%%%%%%%%%%%%%%%%%%%%%%
%%%%%%%%%%%%%%%%%%%%%%%%%%%%%%%%%%%%%%%%%%%%%%%%%%%%%%%%%%%%%%%%%%%
\section{MAGIC observations and data analysis}
\label{sec:2}
MAGIC is a system of two 17\,m diameter imaging atmospheric Cherenkov telescopes 
located on the Canary island of La Palma, which observes the $\gamma$-ray sky 
from $\sim$50\,GeV to more than 50\,TeV. 
The observation of the Perseus cluster started in 2008 with $\sim$24\,hr of 
observation with a single telescope, which did not allow any detection 
\citep{2010ApJ...710..634A}. Since 2009, MAGIC operates in stereoscopic mode, 
which provides much better sensitivity (about a factor 2 above 300\,GeV). 
A major upgrade of the telescopes then occurred during the northern-hemisphere summers of 2011 and 2012 
\citep{2014arXiv1409.6073}.
The improvement in the MAGIC performance after this upgrade is reported by the \cite{2014arXiv1409.5594}.
Here, we combine all the stereoscopic data taken from October 2009 to November 2014.
The observations carried out before the upgrade, from October 2009 to 
February 2011, led to the detection of IC\,310 \citep{2010ApJ...723L.207A} and 
NGC\,1275 \citep{2012A&A...539L...2A}. These data were taken solely during dark 
time at a low zenith angle (from 12$^\circ$ to 36$^\circ$), ensuring the lowest possible
energy threshold ($\sim$100\,GeV at the analysis level). After data-quality selection, this 
first sample consists of 85\,hr of effective time.
The observations carried out after the upgrade from November 2012 to 
November 2014 were performed under heterogeneous conditions, including moonlight 
and large-zenith angles (from 12$^\circ$ to 60$^\circ$), in order to accumulate
the largest amount of data possible and improve the performance at TeV energies. 
After data-quality selection, the second sample consists of 168\,h of effective time.

All the observations were taken in so-called ``wobble'' mode, pointing to alternative
sky positions lying 0.4$^\circ$ from NGC\,1275, which is at the centre of the 
cluster. Most of the pointings were also separated by 0.4$^\circ$ from IC\,310 to allow the 
survey of both AGNs simultaneously. The standard MAGIC Analysis and 
Reconstruction Software (MARS, \citealp{2013ICRC...0773}) was used to analyse the data.  
The results of the 2009--2011 campaign were already reported by the \cite{2011arXiv1111.5544M}.
Here, we re-analyse these data with same calibration, image cleaning, and gamma-hadron separation
as previously and combine them with the more recent data only at the last stage of the 
analysis.
For the 2012--2014 data sample, we use the so-called ``sum image cleaning'' with 
cleaning thresholds that are 33\% higher than the standard cleaning (reported in 
\citealp{2014arXiv1409.5594}) to properly handle the data taken under moonlight 
conditions. This increases the energy threshold from 100\,GeV to about 150\,GeV but does not 
affect the performance at TeV energies. Additionally, only events with more than
150 photo electrons in the camera image of both telescopes are kept. 
Below this cut our simulation of the instrument response with fixed pixel discriminator thresholds does not correctly 
describe the actual MAGIC trigger, which uses adaptive thresholds, during Moon observation.

\begin{figure*}[ht!]
\centering
\includegraphics[width=0.45\textwidth]{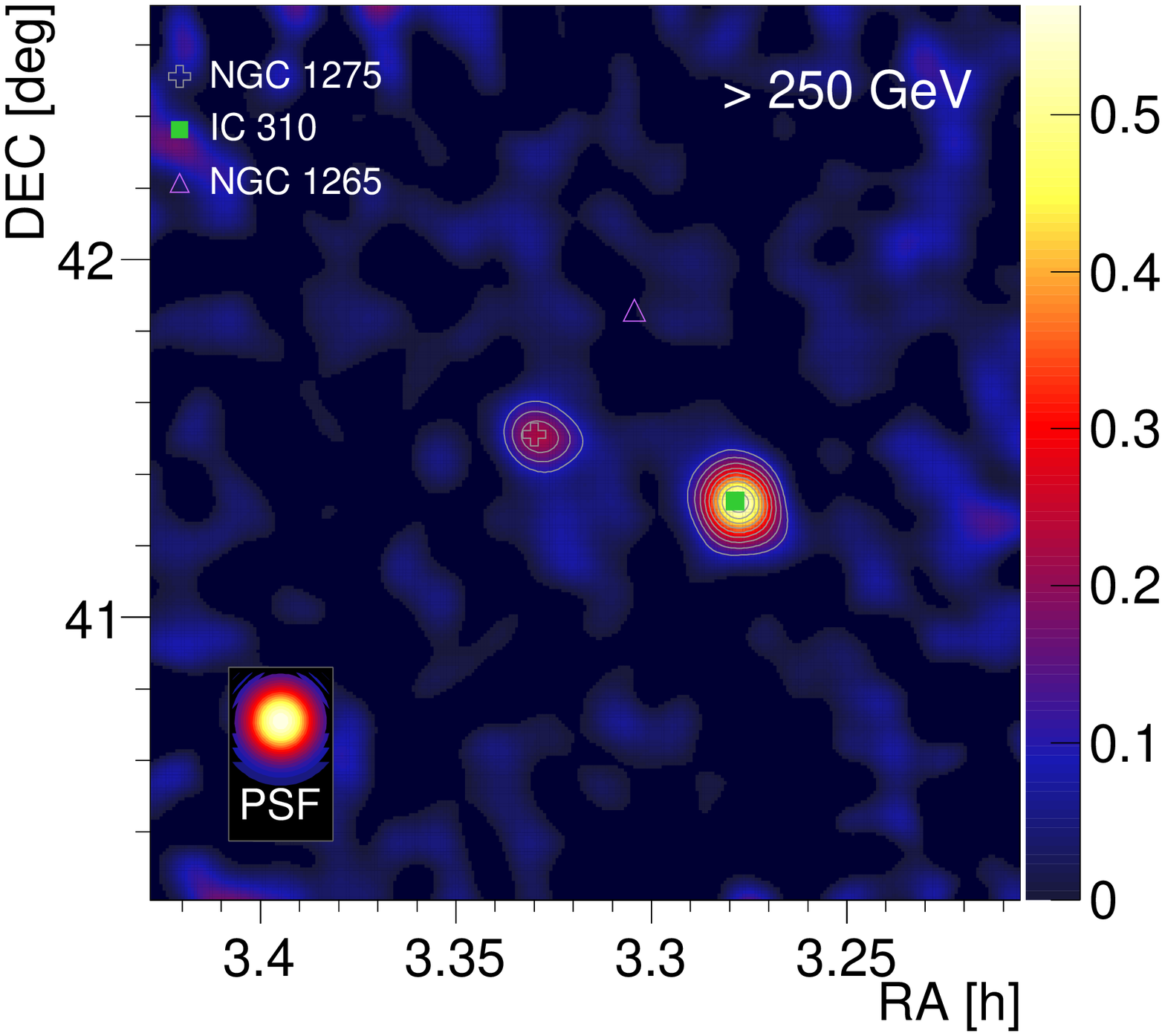}
\includegraphics[width=0.45\textwidth]{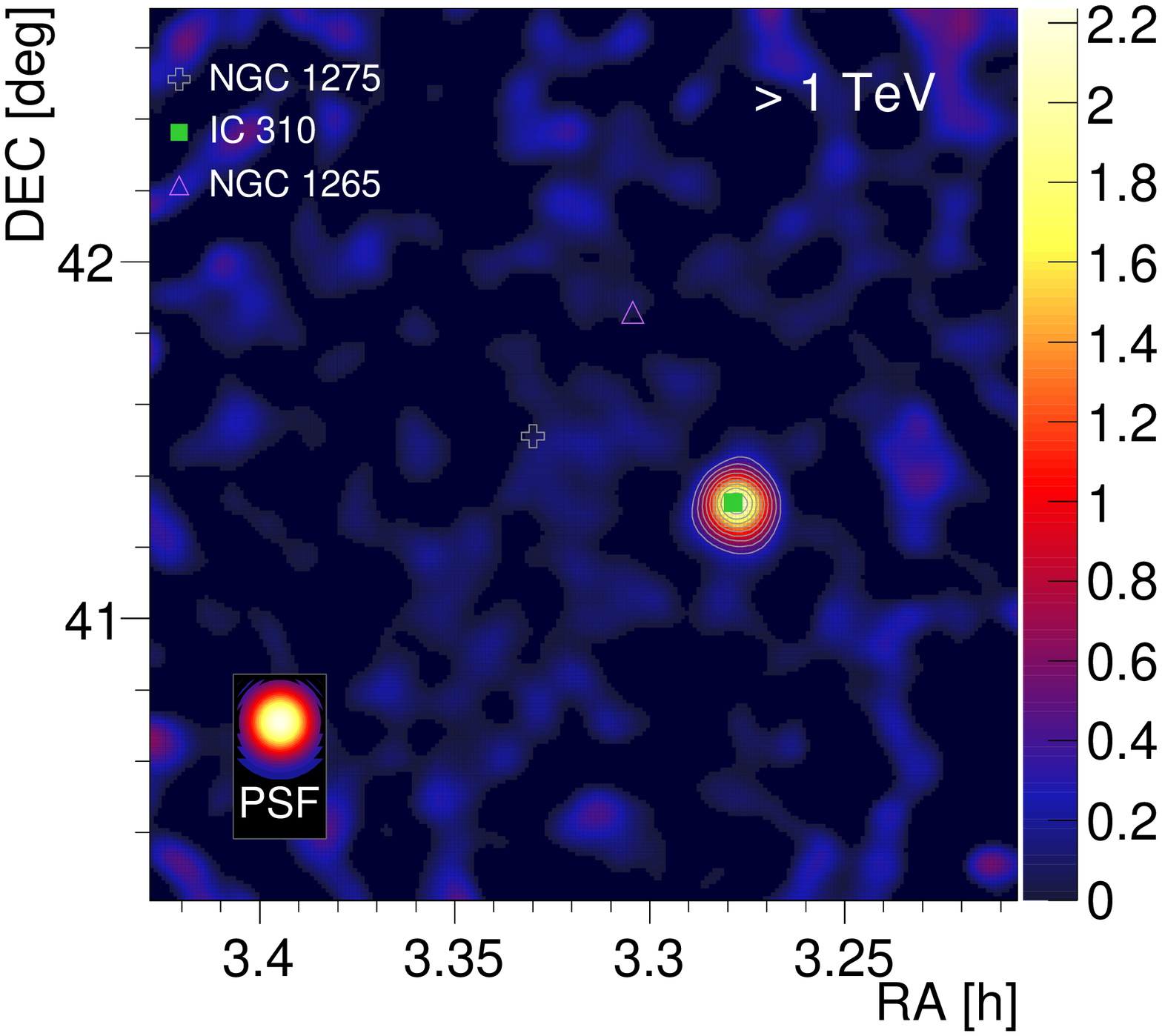}
\includegraphics[width=0.45\textwidth]{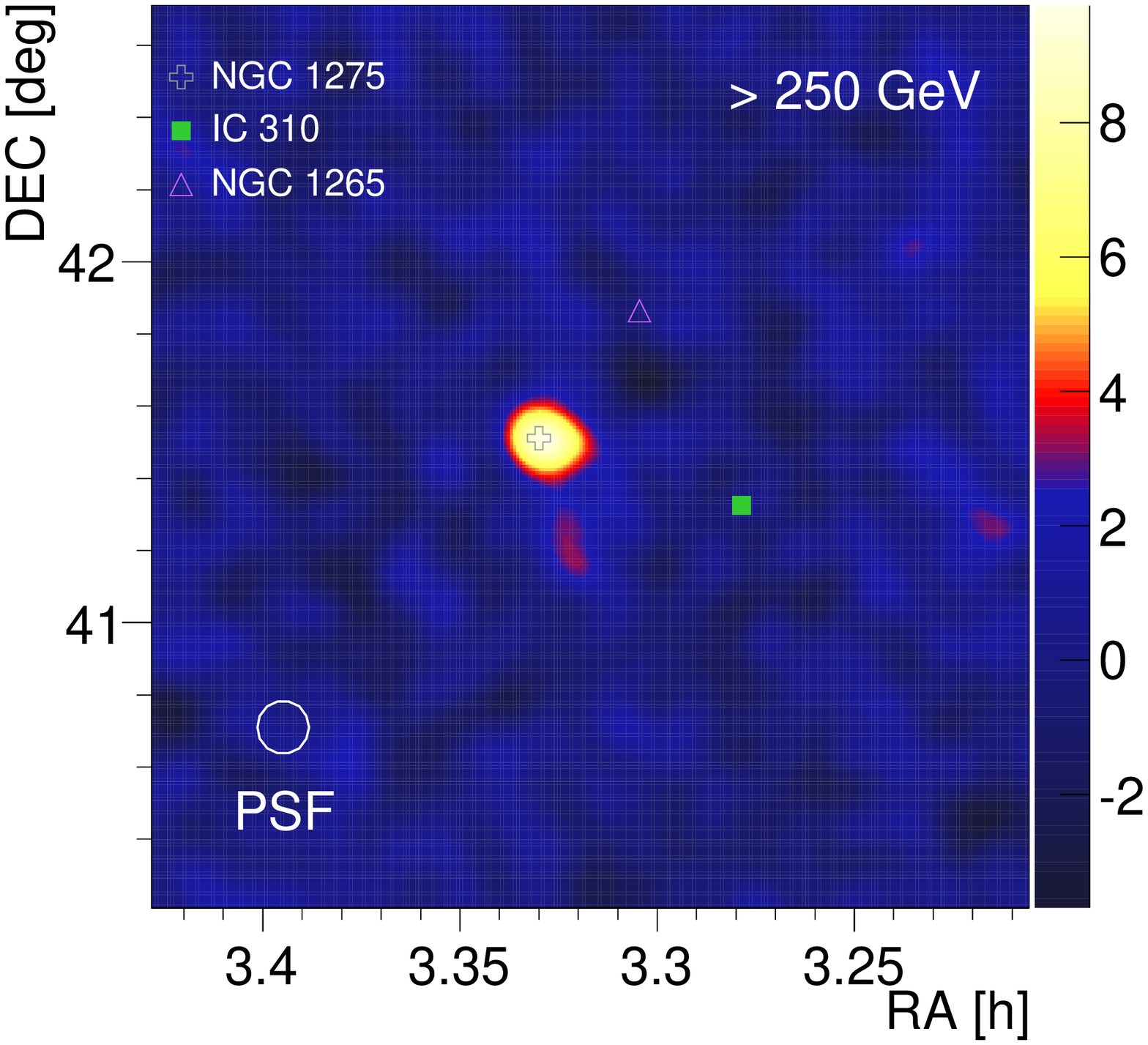}
\includegraphics[width=0.45\textwidth]{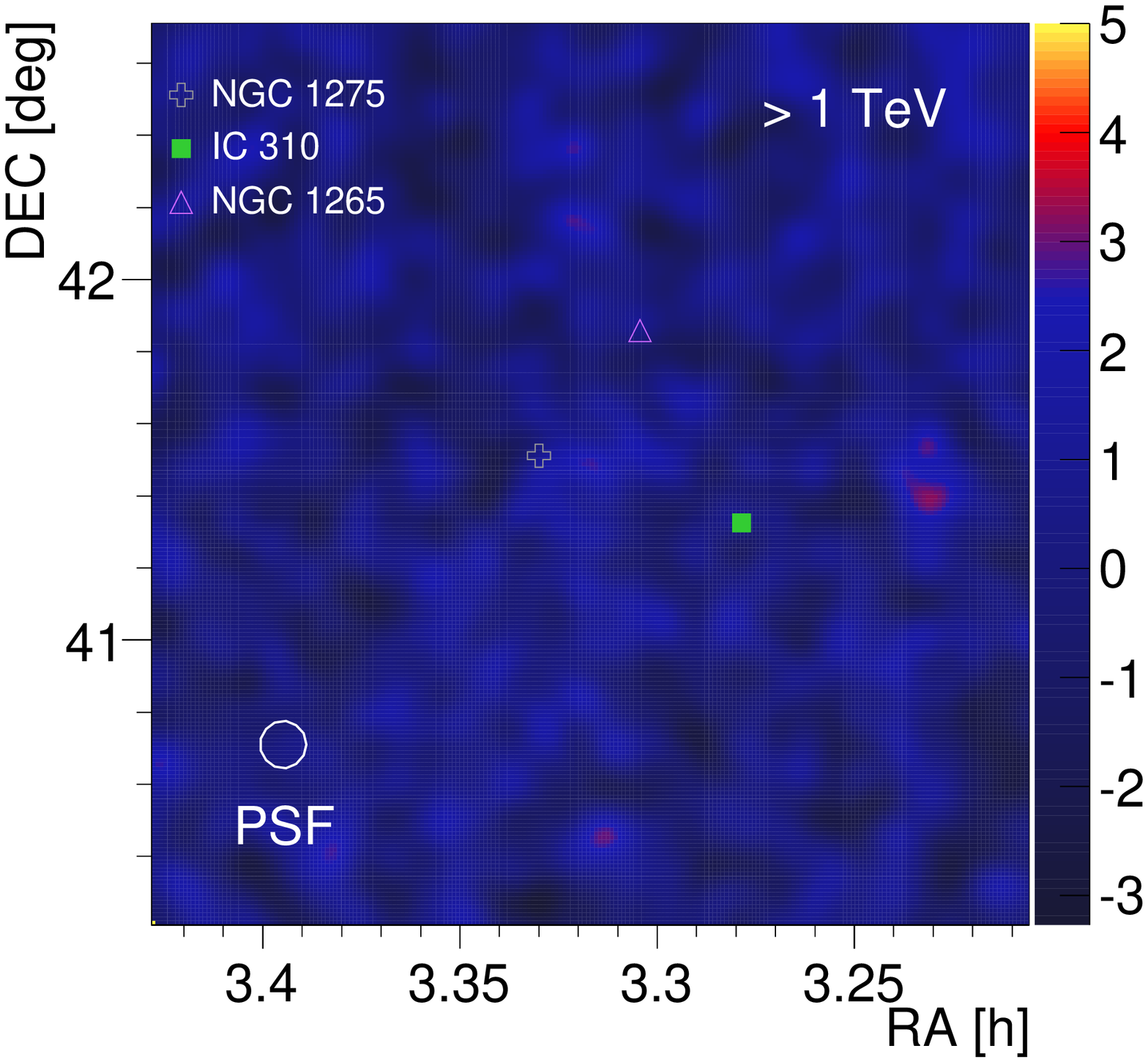}
\caption{Perseus cluster sky maps for an energy threshold of 250\,GeV (left-hand 
panels) and 1\,TeV (right-hand panels) obtained from 253\,hr of MAGIC 
observation. Upper panels show the relative flux (colour code, expressed in 
signal-to-background ratio) and the excess significance (contour lines starting 
from 4~$\sigma$ with steps of 2~$\sigma$). Lower panels show the significance 
maps where the signal from IC\,310 is subtracted (see text). Symbols 
indicate the positions of the three brightest radio galaxies of the cluster.}
\label{fig:SkyMaps}
\end{figure*}

The gamma-hadron separation is performed by the standard MAGIC method using the
Random Forest algorithm \citep{2008NIMPA.588..424A}.
The remaining background from the CR-induced air showers is estimated
from background control regions (OFF regions) lying at 0.4$^\circ$ from the camera centre.
To prevent contamination by the strong IC\,310 signal, the OFF 
regions closer than 0.4$^\circ$ to IC\,310 have been excluded.
Our analysis assumes different source extensions that only differ on the signal-region radius,
$\theta_{\rmn{cut}}$, and the number of OFF regions.
For the point-like source analysis, an average of 5 OFF regions in the field of view are used. In 
the case of extended source analysis, only the most distant OFF region, lying 
0.8$^\circ$ from NGC\,1275, is used. 

%%%%%%%%%%%%%%%%%%%%%%%%%%%%%%%%%%%%%%%%%%%%%%%%%%%%%%%%%%%%%%%%%%%
%%%%%%%%%%%%%%%%%%%%%%%%%%%%%%%%%%%%%%%%%%%%%%%%%%%%%%%%%%%%%%%%%%%
\section{MAGIC results}
\label{sec:3}
\begin{figure}[t!]
\centering
\includegraphics[width=0.5\textwidth]{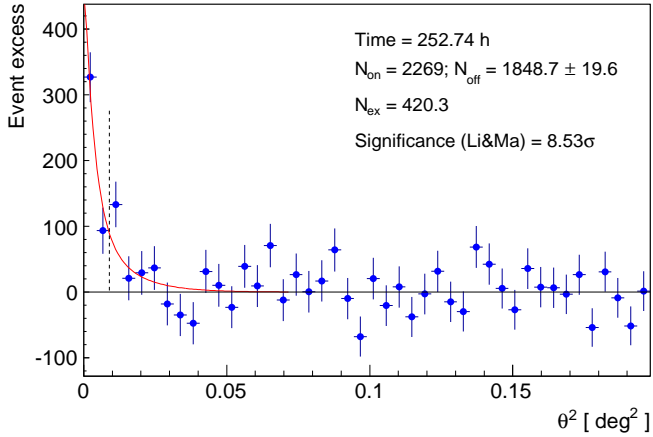}
\caption{Excess signal above 250\,GeV as a function of the squared distance to 
NGC\,1275 ($\theta^2$). The solid red line is the MAGIC PSF fit ,
and the vertical dashed line shows the optimal $\theta^2$ cut used for a point-like source detection.}
\label{fig:NGC1275_theta_plot}
\end{figure}

The 253\,hr of stereo observation from different 
periods were combined to provide the deepest view of the Perseus cluster at 
very high energy. The upper left- and right-hand panels of Figure~\ref{fig:SkyMaps} 
show the relative-flux (i.e., signal-to-background ratio) sky maps for an energy 
threshold of 250\,GeV and 1\,TeV, respectively.
A clear signal is detected from the two previously discovered AGNs.
The bright and hard source IC\,310 is visible in both maps with a high 
significance, and it could mask smaller signals. To search for weak emissions, 
we modelled the point-like emission from IC\,310 as an additional background component. The lower panels of 
Figure~\ref{fig:SkyMaps} show the resulting significance sky maps. Above 
250\,GeV, the emission from NGC\,1275 is detected with a significance of 
8.5$\sigma$ and a signal-to-background ratio greater than 20\%. Above 1\,TeV, however, no 
source other than IC\,310 is detected. 
Figure~\ref{fig:NGC1275_theta_plot} compares the excess ($\gamma$-ray) event 
distribution above 250\,GeV as a function of the squared distance from NGC\,1275 
($\theta^2$) with the MAGIC point spread function (PSF) obtained from 
contemporaneous Crab nebula data at similar zenith angles. In this $\theta^2$ plot, the PSF is described by a double exponential
(corresponding to a double Gaussian in a 2D map), which matches the excess shape of the Crab nebula\footnote{The
Crab nebula has a much harder spectrum than NGC\,1275, so the PSF for NGC\,1275 could be slightly larger.
The PSF shown in Figure~\ref{fig:NGC1275_theta_plot} has been normalised to fit the NGC\,1275 signal. It is just illustrative and
no quantitative result on the intrinsic extension can be derived.} well \citep{2014arXiv1409.5594}.
The shape of the detected signal is in perfect agreement with a point-like object such as the one expected for an AGN.
At both energies, there is no sign of diffuse $\gamma$-ray structures inside the cluster.

\begin{figure}[t!]
\centering
\includegraphics[width=0.5\textwidth]{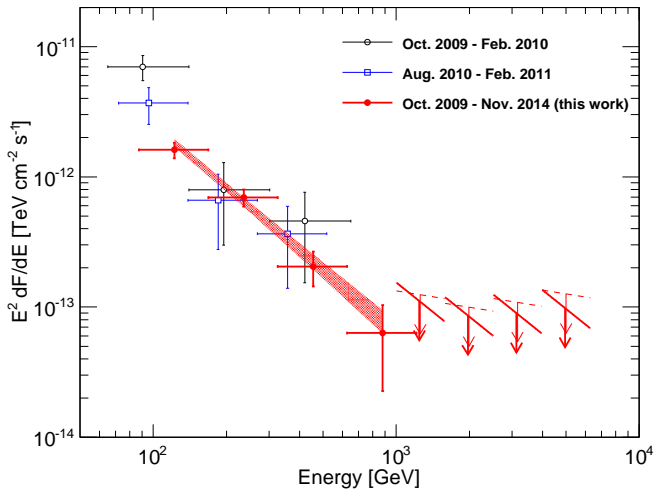}
\caption{Spectral energy distribution of NGC\,1275 averaged over different 
periods with the newest results in red ($\sim$250\,hr). Arrows indicate the 95\%-c.l. differential flux upper limits (5 
bins/decade) for a power-law spectrum with a photon index of $\Gamma=3.5$ (thick 
solid lines) and $\Gamma=2.3$ (thin and dashed lines), respectively.}
\label{fig:NGC1275_sed}
\end{figure}

The average energy spectrum of NGC\,1275 obtained with the full data set (August 
2009 -- November 2014) is shown in Figure~\ref{fig:NGC1275_sed}, together with 
the previous results from the first two years of observation \citep{2014A&A...564A...5A}.
The new spectrum starts at higher energy because the data include moonlight and 
large-zenith-angle observations. With increased photon statistics, we get
better precision and reach higher energies up to 880\,GeV.
This last data point is only marginally significant ($\sim$2$\sigma$)
and in agreement with the upper limits discussed later.
The spectrum between 90\,GeV and 1200\,GeV can be described well by a simple 
power law\footnote{Power-law fit obtained with the forward-unfolding method over 7 reconstructed-energy bins ($\chi^{2}/n_{\rmn{dof}}=2.4/5$).}
\begin{equation}
\frac{\mbox{d}F}{\mbox{d}E} = f_{0} 
\left(\frac{E}{\mathrm{200\,GeV}}\right)^{-\Gamma},
\end{equation}
with a photon index $\Gamma=3.6\pm0.2_{\rmn{stat}}\pm0.2_{\rmn{syst}}$
and a normalisation constant at 200\,GeV of $f_{0} =(2.1 \pm 0.2_{\rmn{stat}} \pm 
0.3_{\rmn{syst}}) \times 10^{-11} \mathrm{cm^{-2} s^{-1} TeV^{-1}}$. 
The systematic errors on the flux normalisation and spectral slope 
take the signal-to-noise ratio into account as explained in \cite{2014arXiv1409.5594}.
Additionally, the uncertainty on the energy scale is 15\%.

The average NGC\,1275 spectrum from August 2009 to November 2014 is
in good agreement with previous measurements \citep{2014A&A...564A...5A}.
The transition between a flat spectrum, $\Gamma\simeq2$, measured by \textit{Fermi}-LAT in 0.1--10\,GeV range,
and the soft spectrum observed above 100\,GeV is confirmed to be smooth, and
better fitted by a log parabola or a broken power law compared to a cut-off 
hypothesis (see discussion in \citealp{2014A&A...564A...5A}).
The AGN-physics interpretation is not substantially modified with respect to our previous results.

Figure~\ref{fig:NGC1275_sed} also reports the differential flux upper limits of four energy bins above 1\,TeV for 
a point-like source with a soft power-law spectrum ($\Gamma=3.5$) as measured in the $0.1-1$\,TeV range,
and for a hard spectral index ($\Gamma=2.3$), as approximately expected from the CR-induced signal (see next section).
Upper limits are calculated using the method of \cite{2005NIMPA.551..493R} 
for a confidence level (c.l.) of 95\% and with a total systematic uncertainty of 30\%.
These results assume a point-like emission, while the CR-induced signal 
should be spatially extended. Upper limits on such diffuse emission depend on 
the assumption of the surface brightness shape. In the next section, we discuss several 
CR-induced emission models and report the associated flux upper limits.

The non-detection of NGC\,1265 in Figure~\ref{fig:SkyMaps} allows us to derive flux upper limits for this radio galaxy. Also,
$\gamma$ rays could be expected from the central AGN or from the bowshock \citep{1998SijbringBruyn} as
speculated earlier for IC\,310 before the flux variability was established, confirming the AGN nature of the emission \citep{2010A&A...519L...6N}.
The NGC\,1265 position is slightly off-centred, and the exposure at this position is about
20\% lower than for NGC\,1275. Assuming a point-like source with the same spectral shape as NGC\,1275
(a power law with photo index $\Gamma=3.6$), the upper limit of the integral flux above 250\,GeV is estimated
to be $5.6\times10^{-13}$\,cm$^{-2}$s$^{-1}$. 

%%%%%%%%%%%%%%%%%%%%%%%%%%%%%%%%%%%%%%%%%%%%%%%%%%%%%%%%%%%%%%%%%%%
%%%%%%%%%%%%%%%%%%%%%%%%%%%%%%%%%%%%%%%%%%%%%%%%%%%%%%%%%%%%%%%%%%%
\section{Search for cosmic-ray induced emission}
\label{sec:4}
As discussed in Section~\ref{sec:3}, the $\gamma$-ray emission from the central 
galaxy NGC\,1275 is consistent with a point-like source, and no diffuse component 
seems to be present. The measured flux around 100\,GeV is much larger than what is 
expected from the CR-induced emission, and it could be outshining the latter.
However, the NGC\,1275 AGN spectrum is very steep, and no signal is detected 
above about 1\,TeV. The CR-induced emission should be harder with no 
intrinsic cut-off in the MAGIC energy range.\footnote{A cut-off is expected at PeV energies owing to the escape of high-energy CR protons 
that are no longer confined in the cluster volume (see, e.g., \citealp{1996SSRv...75..279V,1997ApJ...487..529B,2010MNRAS.409..449P}).}
Therefore, we use the energies above 1\,TeV to search for the possible diffuse CR-induced component. 

In this section, we discuss the expected energy spectrum and spatial distribution of the CR-induced $\gamma$-ray signal
as it should be observed by MAGIC, when taking the absorption during the travel to Earth and the instrument response into account.
Then, we present the optimisation of our analysis to detect such emissions and the derived flux upper limits.
 
%%%%%%%%%%%%%%%%%%%%%%%%%%% 
\subsection{Spectrum expected on Earth}
\label{sec:4.1}
The hydrodynamical simulations of \cite{2010MNRAS.409..449P} suggest a universal
CR-proton momentum spectrum $p^{-\alpha}$ with $\alpha \approx 2.2$ at the 
energies of interest here.
The very high-energy $\gamma$-ray spectrum induced by pion decays from CR-ICM interactions should have 
approximately the same spectral index as the CR spectrum.
However, these $\gamma$ rays can interact by pair production with the extragalactic background light (EBL)
during their travel to Earth, reducing the observed flux.
To convert the intrinsic spectrum from the Perseus cluster at $z=0.018$
into the spectrum observed on Earth, we use the EBL model of \cite{Dominguez2011}.
Below 300\,GeV, the EBL absorption is negligible (<4\%). 
Between 300\,GeV and 10\,TeV, the effect of the absorption can be approximated 
by an increase (softening) in the power-law index of about $0.13$ and a reduction of the 
differential-flux normalisation at 1\,TeV of 17\%. Above 10\,TeV, the absorption 
increases dramatically. A source with an intrinsic power-law spectrum index of $\alpha=2.2$ would appear above 300\,GeV
as a power law with an index $\Gamma=2.33$ and a cut-off above 10\,TeV. About 20\% of the flux above 1\,TeV and 60\% above 10\,TeV 
is absorbed during the travel to Earth.

There are three different scenarios for the fate of the ultra-relativistic electron-positron pairs
that are produced by $\gamma$-ray$-$EBL interactions. 
Each possibility predicts a different signal for the energy range of interest here.

\begin{enumerate}

\item The pairs can interact with CMB photons that are Compton up-scattered in energy. 
As a result, an initial TeV $\gamma$-ray is reprocessed to GeV energies. 
If the initial $\gamma$ ray has a much higher energy, this process can be repeated to produce 
the so-called inverse Compton cascade (ICC) effect \citep[see, e.g.,][]{NeronovVovk2010}. 
To first approximation, the ICC emission into the energy range of interest here, $\sim$1--10\,TeV, 
is induced by primary $\gamma$ rays from the energy range $\sim$30--100\,TeV. Assuming the most 
favourable case, when all the absorbed energy is re-emitted, the ICC emission from a power-law 
spectrum with $\alpha = 2.2$ would represent about 30\% of the intrinsic emission of the $\sim$1--10\,TeV range.\\

\item The compact ICC signal could be diluted by the presence of extragalactic magnetic field (EGMF), which deflects 
the electron and positron tracks and, consequently, induces angular dispersion, as well as a time delay of the signal
\citep[see, e.g.,][]{NeronovVovk2010,2010MNRAS.406L..70T,2011MNRAS.414.3566T,2011ApJ...733L..21D,
2011A&A...529A.144T,2011ApJ...727L...4D,2012ApJ...744L...7T}. 
Being constant, the CR-induced emission is not affected by the time delay.
According to the approximation of \citet{NeronovVovk2010}, the ICC emission induced 
by 50\,TeV $\gamma$ rays from 78\,Mpc should have a spatial extension of the order of $0.1^\circ$ for an EGMF $\approx 10^{-14}$\,G.
For a weaker EGMF, the spatial distribution of the ICC would be almost the same as the intrinsic 
CR-induced emission, and one could expect up to 30\% higher signal in the $\sim$1--10\,TeV range (assuming the intrinsic 
power-law spectrum extends to 100\,TeV). For an EGMF $\gg 10^{-14}$\,G, however, the ICC would be 
diluted in the diffuse extragalactic $\gamma$-ray background. Thus, depending on the EGMF level, the $\gamma$-ray loss
due to the EBL absorption in the $\sim$1--10\,TeV range could be compensated for by the ICC emission, up to a full 
compensation for a very low EGMF.\\ 

\item A competing mechanism exists that could modify the evolution of the pairs on a faster time scale than the ICC,
namely powerful plasma instabilities driven by the anisotropy of the ultra-relativistic pair beams \citep{2012ApJ...752...22B,2012ApJ...758..101S,2012ApJ...758..102S,
2012ApJ...752...23C,2012ApJ...752...24P,2013ApJ...770...54M,2013ApJ...777...49S,2014ApJ...787...49S,2014ApJ...783...96S,2014ApJ...797..110C}.
This picture is interesting because it can match the observed extragalactic $\gamma$-ray background spectrum above 3\,GeV and flux distributions of TeV blazars simultaneously, 
using a unified model of AGN evolution \citep{2014ApJ...790..137B,2014ApJ...796...12B}. 
In contrast to the set-up studied by \citet{2012ApJ...752...22B}, here, we have to compare the oblique instability 
growth rate, $\Gamma_{\mathrm{M,k}}\propto n_{\mathrm{beam}}$ (where $n_{\mathrm{beam}}$ is density of beam pairs), 
to the IC cooling rate, $\Gamma_{\mathrm{IC}}$, at a fixed distance to the source. The mean-free-path to pair 
production, $d_{\mathrm{pp}}$, of primary $\gamma$ rays with energy $>$15\,TeV is smaller than $D_\mathrm{lum}=78$\,Mpc, the distance of the 
Perseus cluster. Hence, the density of beam pairs is lower at $D_\mathrm{lum}$ in comparison to $d_{\mathrm{pp}}$.
As the expected cluster luminosity is smaller than the minimum luminosity of $E\,L_{E,\mathrm{min}} \approx 10^{42}\,\mbox{erg s}^{-1}$
(at $z\approx0$ and $d_{\mathrm{pp}}$ from the source) needed for the oblique instability to grow faster than IC cooling rate,
$\Gamma_{\mathrm{M,k}}<\Gamma_{\mathrm{IC}}$ at $d_{\mathrm{pp}}$ (a fortiori at $D_\mathrm{lum}$),
the absorbed 30--100\,TeV $\gamma$-ray flux is very likely reprocessed via ICCs to our energy range of interest.
\end{enumerate}

The EBL effect was neglected in all previous papers on this topic, for which, therefore, the constraints derived
from $\gamma$-ray flux upper limits above $\sim$1\,TeV can be considered as optimistic. 
They correspond roughly to the case of a strong ICC emission (EGMF $<10^{-14}$\,G), which compensates for the EBL absorption.
Here, in contrast, we include EBL absorption in our reference case and assume the most conservative case without ICC emission.
Also, CR propagation effects, such as streaming and diffusion, could cause a softening of the CR spectrum.
In the following we do not consider this possibility because of the rather uncertain modelling at this stage (e.g., \citealp{2013arXiv1303.4746W}).

%%%%%%%%%%%%%%%%%%%%%%%%%%% 
\subsection{Flux upper limits for different spatial distributions}
\label{sec:4.2}
\begin{figure}[t!]
\centering
\includegraphics[width=0.5\textwidth]{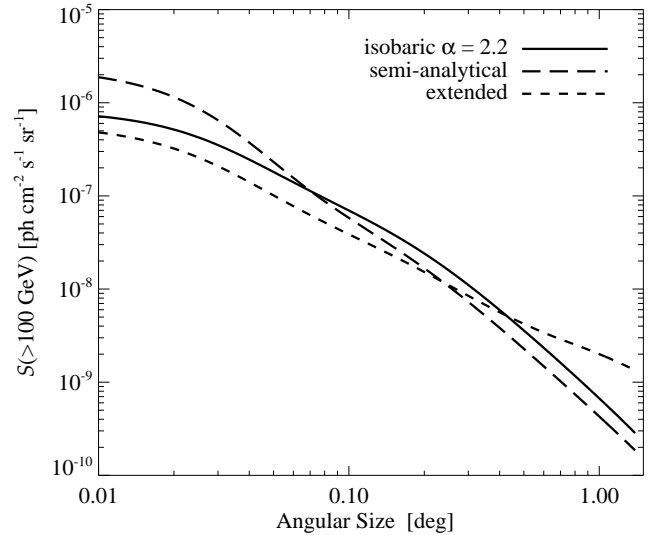}
\caption{Surface brightness profiles, above 100\,GeV, of the three tested 
spatial templates for the CR-induced emission in Perseus: \emph{isobaric} with 
$\alpha=2.2$, \emph{semi-analytical}, and \emph{extended}. 
The normalisation of the \emph{isobaric} model is set to respect our 
previous upper limits \citep{2011arXiv1111.5544M}, while the normalisations of 
the \emph{semi-analytical} and \emph{extended} models are as from 
\cite{2010MNRAS.409..449P} and \cite{2012arXiv1207.6410Z}, respectively. See 
main text for details.}
\label{fig:TheoreticalProfiles}
\end{figure}

\begin{figure}[t!]
\centering
\includegraphics[width=0.48\textwidth]{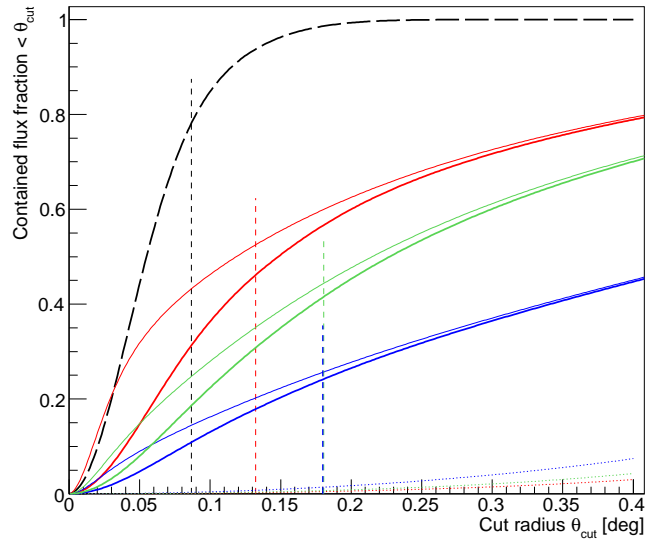}
\caption{Cumulative fraction of signal within a given radius for different 
models: \emph{point-like} in black (long-dashed line), 
\emph{isobaric} in green, \emph{semi-analytical} in red, and \emph{extended} in 
blue. The distribution of the \emph{point-like} model follows the MAGIC PSF 
above 630\,GeV. The thin and thick coloured solid lines represent the real and 
reconstructed (i.e., smeared by the PSF) signal fractions, respectively.
The dotted lines represent the fraction of these signals contained in the 
reference background region. The vertical dashed lines correspond to the used
optimum $\theta_{\rmn{cut}}$ values.}
\label{fig:CumulativeSignal}
\end{figure}

To account for our limited knowledge of the spatial shape of the CR-induced emission, 
we adopted three different models that describe the CR density as a function of the distance from the cluster centre: 
i) the \emph{isobaric} model assuming a constant CR-to-thermal pressure $X_\mathrm{CR} 
= P_\mathrm{CR}/P_\mathrm{th}$ \citep{2004A&A...413...17P}; ii) the 
\emph{semi-analytical} model of \cite{2010MNRAS.409..449P} derived from hydrodynamical 
simulations of clusters; and iii) the \emph{extended}
hadronic model of \cite{2012arXiv1207.6410Z}, where the possibility of CR 
propagation out of the cluster core is considered,
resulting in a significantly flatter CR profile.\footnote{The \emph{extended}
model adopted in this work corresponds to the model with $\gamma_{\rmn{tu}}=3$
in \cite{2012arXiv1207.6410Z}, where $\gamma_{\rmn{tu}}$ is a parameter
that indicates the dominant CR transport mechanism.} 
In Figure~\ref{fig:TheoreticalProfiles}, we show the $\gamma$-ray surface 
brightness above 100\,GeV for the three models. In all cases, fundamental input 
parameters are the Perseus ICM density and temperature as measured in X-rays 
\citep{2003ApJ...590..225C}. 
The fraction of signal expected within a circular region of a given radius 
$\theta$ from the cluster centre is shown in Figure~\ref{fig:CumulativeSignal}, 
together with the MAGIC PSF above 630\,GeV. 

Since the predicted CR-induced signal extension is significantly larger than the 
PSF, the optimal $\theta_{\rmn{cut}}$ used to detect the emission is different than 
for a point-like source. Comparing the predicted signal in the ON region 
($\theta<\theta_{\rmn{cut}}$) with the corresponding background level 
estimated from our data, and taking the expected signal leakage 
inside the OFF region at 0.8$^\circ$ from the cluster centre into account, we optimised the 
$\theta_{\rmn{cut}}$ for each model. In practice, $\theta_{\rmn{cut}}$ is optimal on 
a relatively wide range. The same cut can be used for 
both the \emph{isobaric} and \emph{extended} models, which have very similar optimum $\theta_{\rmn{cut}}$ values.
Since the MAGIC PSF is relatively stable above 630\,GeV \citep{2014arXiv1409.5594}, we use
the same $\theta^2_{\rmn{cut}}$ value for all energies for any given model. The resulting optimum cuts are presented in 
Table~\ref{tab:Theta2Cuts} and shown in Figure~\ref{fig:CumulativeSignal}.

\begin{table}[t!]
\caption{Optimum $\theta^2_{\rmn{cut}}$ and contained flux fractions for different 
models}
\begin{center}
\begin{tabular}{lcccc}
\hline\hline
\phantom{\Big|}
Models & $\theta^2_{\rmn{cut}}$ & flux($<\theta_{\rmn{cut}}$) & flux($<0.15^\circ$)\tablefootmark{a}\\
       & [deg$^2$] & ON~~(OFF) & real~~(recon.) \\
\hline\\[-0.5em]
\emph{point-like} & 0.0075 &  77\% (0.0\%)   & 100\% (96\%) \\
\emph{isobaric}   & 0.0325 &  41\% (0.7\%) & 39\% (35\%) \\
\emph{semi-analytical} & 0.0175 & 46\% (0.2\%) & 55\% (50\%) \\
\emph{extended}   & 0.0325 &  26\% (1.4\%) & 22\% (20\%) \\
\hline
\end{tabular}
\end{center}
{\bf Notes.} \tablefoottext{a}{The flux within 0.15$^\circ$ is used as reference 
for all models.}
\label{tab:Theta2Cuts}
\end{table}%

Concerning the CR-proton spectrum, we adopt for both the \emph{semi-analytical} and \emph{extended} models, the universal
power-law momentum spectrum $p^{-\alpha}$, with $\alpha \approx 2.2$ for the energies of interest here, found in hydrodynamical 
simulations by \cite{2010MNRAS.409..449P}. For the less predictive \emph{isobaric} model, the spectral index is free to vary, and we 
assume a range of values, $2.1\leq\alpha\leq2.5$.
In all cases, we also consider the EBL absorption, which results in a slightly softer observed $\gamma$-ray spectrum,
$\Gamma = \alpha + 0.13$, in the energy range 300\,GeV--10\,TeV.

Table~\ref{tab:ULs} presents the 95\%-c.l. upper limits of the integral flux
between several energy thresholds ($E_{\rmn{th}}$) and 10\,TeV for different spatial models
assuming a power-law spectral index $\Gamma=2.33$.
The upper limits are converted to the corresponding flux contained within the reference radius $0.15^\circ$ and 
the cluster virial radius\footnote{The cluster virial radius is defined 
here with respect to an average density that is 200 times the critical density of the 
Universe.} $R_{200} \simeq 1.4^\circ$ \citep{2002ApJ...567..716R}. The 
conversion factors depend on the surface brightness distributions, also considering 
the signal contained in the OFF region. The latter is particularly important for 
the \emph{extended} CR model, which has a flatter surface brightness 
profile. For the \emph{point-like} assumption, we estimated the background level 
from an average of five OFF regions, while for the CR models we only
used a single 
region at 0.8$^\circ$ from the cluster centre, as mentioned in 
Section~\ref{sec:2}.

\begin{table}[t!]
\caption{Integral flux upper limits between $E_\rmn{th}$ and 10\,TeV assuming 
observed power-law spectrum with an index $\Gamma=2.33$}
\begin{center}
\resizebox{0.48\textwidth}{!}{
\begin{tabular}{lcccccc}
\hline\hline
\phantom{\Big|}
$E_\rmn{th}$ & model & $N_\rmn{ON}$\tablefootmark{a} & $N_\rmn{OFF}$\tablefootmark{a} & 
$\sigma_\rmn{LiMa}$\tablefootmark{b} & $F_\rmn{UL}^{0.15^\circ}$\tablefootmark{c} & 
$F_\rmn{UL}^{1.4^\circ}$\tablefootmark{c}\\
\hline\\[-0.5em]
         &\emph{point-like}    & 332 & 304.1 & 1.4 & 12.2 &  12.2 \\
630\,GeV &\emph{isobaric}      & 1327 & 1256 &1.4 & 24.8 &  65.5 \\
         &\emph{semi-analytic} & 749 & 681 & 1.8 & 26.8 &  49.0 \\
         &\emph{extended}      & 1327 & 1256 & 1.4 & 25.6 &  124. \\
\hline
         &\emph{point-like}    & 159 & 157.5 & 0.1 & 3.84 &  3.84  \\
1.0\,TeV &\emph{isobaric}      & 675 & 652 & 0.6 & 10.7 &  28.3  \\
         &\emph{semi-analytic} & 369 & 352 & 0.6 & 9.77 &  17.9  \\
         &\emph{extended}      & 675 & 652  & 0.6 & 11.1 &  54.0  \\
\hline
         &\emph{point-like}    & 77 & 75.9 & 0.1 & 2.34 &  2.34  \\
1.6\,TeV &\emph{isobaric}      & 321 & 317 & 0.2 & 5.09 &  13.5  \\
         &\emph{semi-analytic} & 169 & 167 & 0.1 & 4.61 &  8.43  \\
         &\emph{extended}      & 321 & 317 & 0.2 & 5.26 &  25.6  \\
\hline
         &\emph{point-like}    & 38 & 37.4 & 0.1 & 1.57 &  1.57  \\
2.5\,TeV &\emph{isobaric}      & 143 & 153 & -0.6 & 2.18 &  5.75  \\
         &\emph{semi-analytic} & 77 & 81 & -0.3 & 2.30 &  4.21  \\
         &\emph{extended}      & 143 & 153 & -0.6 & 2.25 &  11.0  \\
\hline
2.5\,TeV &\emph{isobaric}      & - & - & - & 3.10 &  8.15  \\
for zero &\emph{semi-analytic} & - & - & - & 2.81 &  5.15  \\
excess\tablefootmark{d}&\emph{extended}   & - & - & - & 3.21 &  15.5  \\
\hline
\end{tabular}
}
\end{center}
{\bf Notes.} \tablefoottext{a}{Number of events in the signal ($N_\rmn{ON}$) and 
background ($N_\rmn{OFF}$) regions.}
\tablefoottext{b}{Significance of the measured excess in standard deviations.}
\tablefoottext{c}{95\%-c.l. flux upper limits in units of 10$^{-14}$\,cm$^{-2}$\,s$^{-1}$ 
within a radius of 0.15$^\circ$ and 1.4$^\circ$ from the cluster centre.}
\tablefoottext{d}{For negative measured excess, we provide conservative upper 
limits assuming zero excess.}
\label{tab:ULs}
\end{table}%

The effective area of MAGIC is relatively flat above 630\,GeV, and the integral 
flux upper limits do not depend strongly on the assumed spectral shape. 
Therefore, the upper limits reported in Table~\ref{tab:ULs} are valid, within 2\%,
for an observed spectral index range of $2.1\la\Gamma\la2.6$, which 
corresponds to an intrinsic index range of about $2.0\la\alpha\la2.5$.

%%%%%%%%%%%%%%%%%%%%%%%%%%%%%%%%%%%%%%%%%%%%%%%%%%%%%%%%%%%%%%%%%%%
%%%%%%%%%%%%%%%%%%%%%%%%%%%%%%%%%%%%%%%%%%%%%%%%%%%%%%%%%%%%%%%%%%%
\section{Interpretation and discussion}
\label{sec:5}
The flux upper limits reported in Section~\ref{sec:4} allow to constrain the CR content in the Perseus cluster. 
We discuss the implications for the CR-to-thermal pressure for the three adopted models: \emph{isobaric}, \emph{semi-analytical,} 
and \emph{extended}. Additionally, for the \emph{semi-analytical} model of \cite{2010MNRAS.409..449P}, our constraints
can be translated into constraints on the CR acceleration efficiency at structure formation shocks. 
Finally, assuming that the Perseus radio mini-halo is induced by secondary electrons (i.e., assuming that the hadronic model for 
the radio mini-halo is valid), we discuss the expected minimum $\gamma$-ray flux and derive constraints on the cluster magnetic field. 

%%%%%%%%%%%%%%%%%%%%%%%%%%%%%%%%%%%%%%%%%%%%%%%%%%%%%%%%%%%%%%%%%%%
\subsection{CR acceleration efficiency}
A major uncertainty in modelling CR physics in clusters of galaxies is the CR-acceleration efficiency, i.e., the percentage of the energy dissipated
in structure formation shocks, which goes into particle acceleration. We define
the CR-acceleration efficiency as in \cite{2010MNRAS.409..449P}as the ratio of the energy density 
in freshly injected CRs to the total dissipated energy density in the downstream region of a shock.
We emphasise that it is particularly difficult to put meaningful constraints on this fundamental quantity 
because of the lack of a general theory of particle acceleration. In fact, our current knowledge of the dependence 
of the acceleration efficiency with the shock Mach number is mainly phenomenological and numerical. Nevertheless, 
we assess the impact of our upper limits in the context of the state-of-the-art model by \cite{2010MNRAS.409..449P} 
and caution that our results are also subject to the limitations stated in that work.

\begin{figure*}[ht!]
\centering
\includegraphics[width=0.49\textwidth]{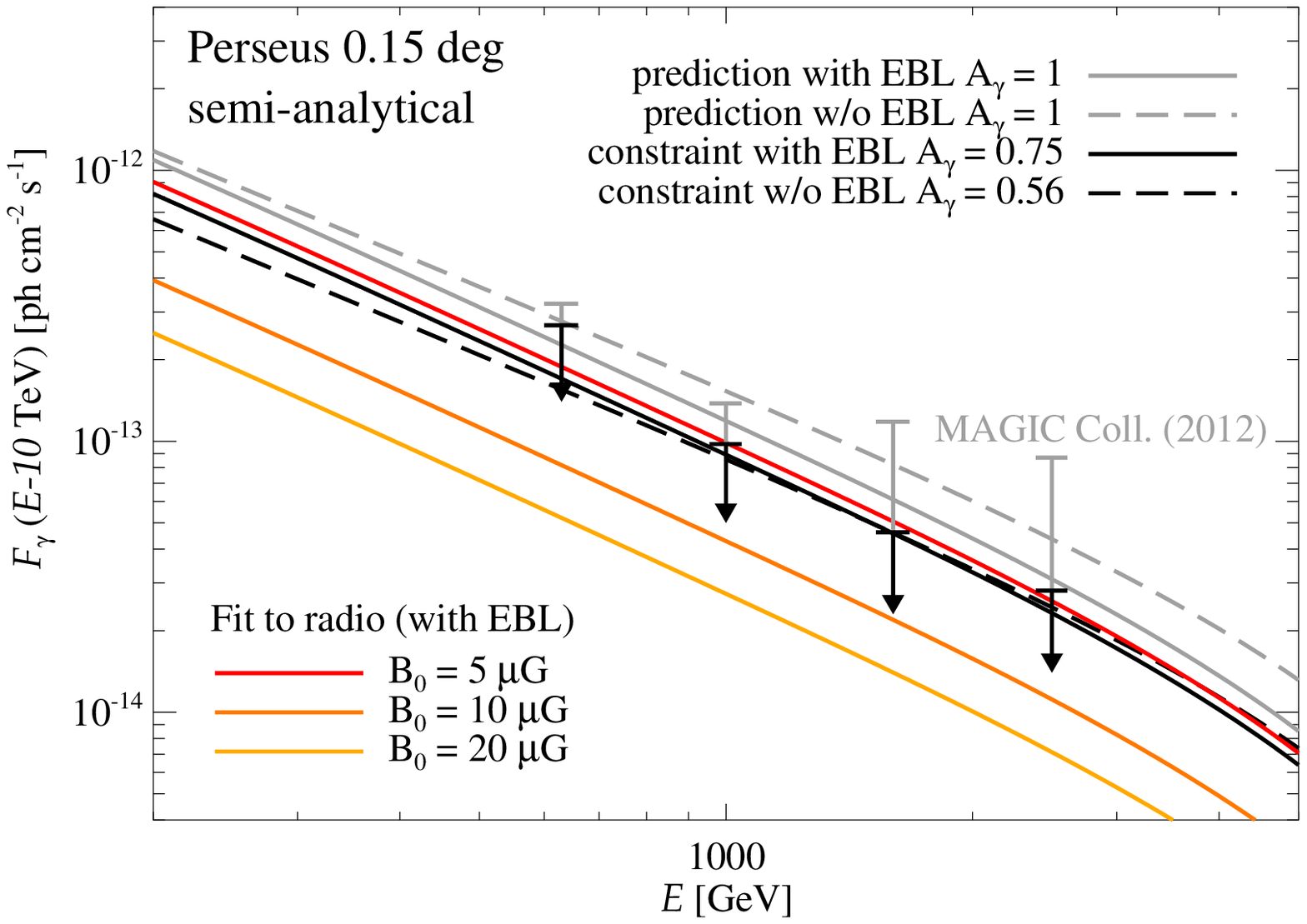}
\includegraphics[width=0.49\textwidth]{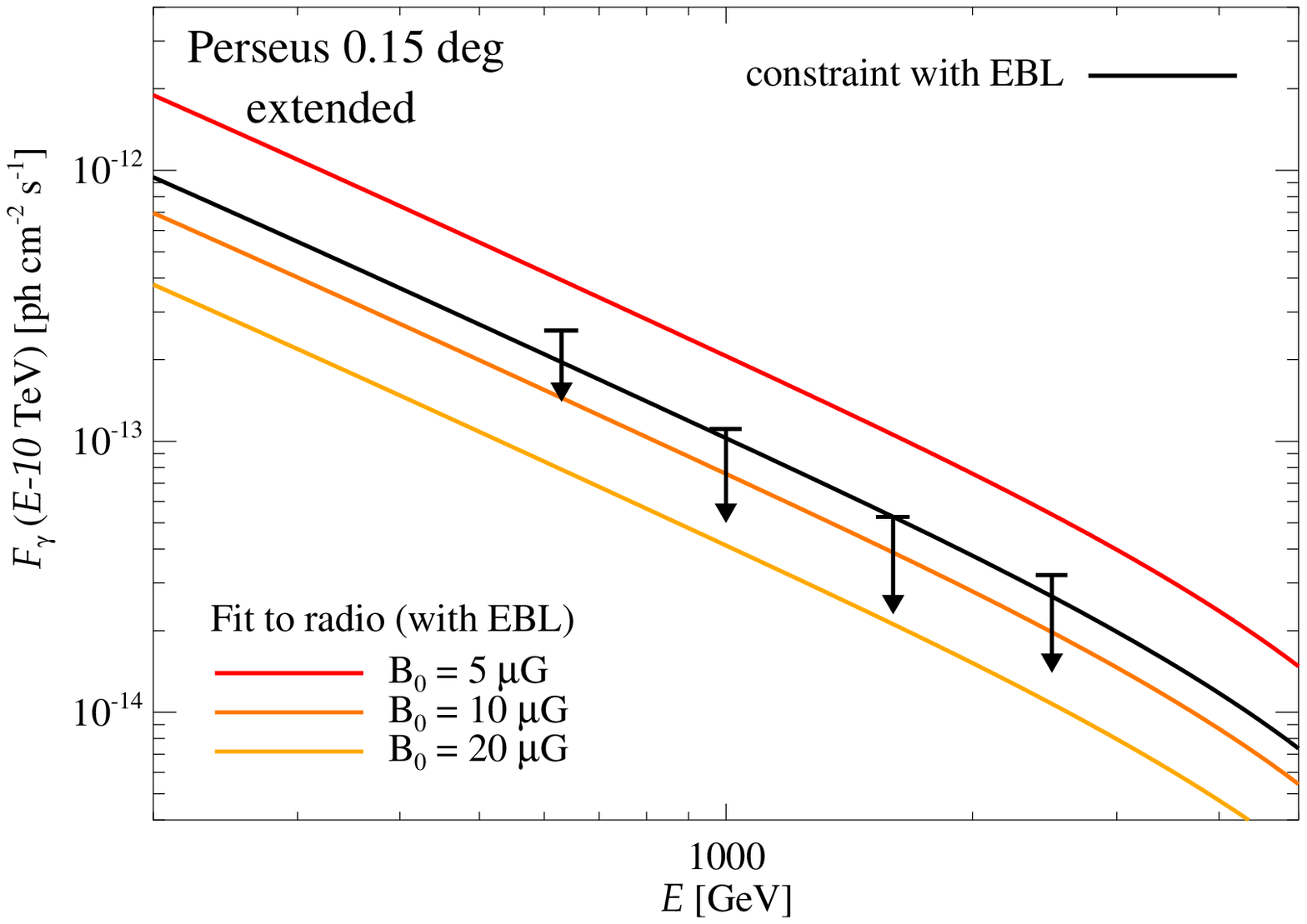}
\caption{
Expected integral flux from the Perseus cluster core within $0.15^\circ$ and the associated 95\%-c.l.
MAGIC upper limits for the \emph{semi-analytical} model ({\bf \emph{Left}}) and the \emph{extended} model ({\bf \emph{Right}}).
For the \emph{semi-analytical} model, we show the spectrum predicted by \cite{2010MNRAS.409..449P}
($A_\gamma = 1$, see text) and the one constrained by our upper limits for both cases with (solid lines) and without (dashed lines) EBL absorption. 
Arrows in light grey show our previous flux upper limits \citep{2011arXiv1111.5544M}.
For both the \emph{semi-analytical} and \emph{extended} models, the EBL-absorbed spectra 
obtained by assuming that the Perseus radio mini-halo has a hadronic origin and 
fixing a central magnetic field value $B_{0} = 5, 10,$ and $20$\,$\mu$G are shown in colours. 
We used as reference the radio surface brightness radial profile of the Perseus 
mini-halo at 1.4\,GHz from \cite{1990MNRAS.246..477P}. Fluxes are integrated up to 10\,TeV.}
\label{fig:ConstrainedSpectra}
\end{figure*}

Our \emph{semi-analytical} model is based on predictions by 
\cite{2010MNRAS.409..449P}, which are derived from hydrodynamical simulations
of galaxy clusters assuming a maximum CR-proton acceleration efficiency 
$\zeta_\mathrm{p, max}= 50$\%. We parametrize $\zeta_\mathrm{p, max}$
with a flux multiplier $A_\gamma$, which is equal to $1$ for $\zeta_\mathrm{p, max}= 50$\%,
and we assume a linear scaling between $A_\gamma$ and $\zeta_\mathrm{p, max}$ \citep{2013arXiv1308.5654T}.
The hadronic-induced emission is proportional to $A_\gamma$, which provides the 
overall normalisation of the CR distribution. In the
left-hand panel of Figure~\ref{fig:ConstrainedSpectra}, we show the integral $\gamma$-ray fluxes
predicted by \cite{2010MNRAS.409..449P} for the Perseus cluster with $A_\gamma=1$ within $0.15^\circ$ from the centre
and the highest fluxes allowed by our observational upper limits. The value of 
$A_\gamma$ is constrained to be $\leq 0.56$ when neglecting the EBL absorption
and $A_\gamma \leq 0.75$ when including the EBL absorption from \cite{Dominguez2011}.
This corresponds to $\zeta_\mathrm{p, max} \leq$\,28\% and $\leq$\,37\%, respectively.

Compared to our previous paper \citep{2011arXiv1111.5544M}, we accumulated three times more data
and derived upper limits that are significantly lower (except for the upper limit at $E>630$\,GeV
that suffers contamination from the NGC\,1275 AGN).
Our previous constraint, $A_\gamma \leq 0.8$, derived from the upper limit at $E>1$\,TeV, was underestimated
because both the EBL absorption and the signal leakage in the OFF regions were neglected.
These effects relax the constraint by about 25\%.
As discussed in Section~\ref{sec:4.1}, this loss can be compensated by ICC emission.
Moreover, the uncertainty on the EBL and ICC effects stays within the 30\% systematics considered in deriving the upper limits.
Here, considering the most conservative case, we can constrain $\zeta_\mathrm{p, max}\leq 37$\%, which is
even slightly below our previous results. We note that \emph{Fermi}-LAT observations suggest values 
of $\zeta_\mathrm{p, max} \lesssim 15$\% for the case of the merging Coma cluster \citep{2013arXiv1308.5654T,2014MNRAS.440..663Z},
although all these results assume a negligible active CR transport, hence represent model-dependent upper limits.

We are able to constrain the CR acceleration efficiency only in the context of 
the hydrodynamical simulations of \cite{2010MNRAS.409..449P} because a full modelling of the formation history of a galaxy cluster and of the CR acceleration 
at the corresponding structure formation shocks is needed for this. Therefore, our constraint on $\zeta_\mathrm{p, max}$ is strictly 
valid only in this context, which assumes in particular 
i) no CR transport relative to the plasma rest frame, and
ii) a simplified model for CR acceleration in which $\zeta_\mathrm{p,max}$ rises steeply for shocks with weak Mach numbers $M$ and already saturates at $M \gtrsim 3$ \citep{2007A&A...473...41E}.
We assess the first assumption by adopting the \emph{extended} model, shown in the right-hand panel of Figure~\ref{fig:ConstrainedSpectra}, 
which was constructed independently of $\zeta_\mathrm{p, max}$ to match the radio emission from the Perseus mini-halo 
(\citealp{2012arXiv1207.6410Z}; see Section~5.3). 
Concerning the second assumption, non-linear diffusive shock acceleration models \citep{kang11,2013ApJ...764...95K}
predict a slower rise in $\zeta_\mathrm{p, max}$ with increasing $M$, which then saturates at higher Mach numbers with respect to \cite{2007A&A...473...41E}.
However, Cherenkov telescopes probe the high-momentum part of the CR-proton spectrum,
which is generated by intermediate-strength shocks where the acceleration efficiency is close to 
saturation for all models. 

Our result on the maximum CR acceleration efficiency is a useful proxy for calibrating 
the total expected emission in the context of the adopted \emph{semi-analytical} model, but
as stressed above, it should be taken with a grain of salt as general constraint in the context of
CR acceleration at shocks. Additionally, the latest supernova remnant observations tend to 
suggest values below 27\% \citep{2013MNRAS.435..910H}, with a commonly accepted value around 
10\% to be able to explain Galactic CRs (e.g., \citealp{2012A&A...538A..81M}). The 
acceleration conditions at structure formation shock may be different from those at supernova remnants, 
complicating any direct comparison.

%A promising new way to investigate CR acceleration efficiencies, and to potentially explain 
%the current non detection of $\gamma$ rays from clusters, comes from particle-in-cell and hybrid 
%simulations of shock acceleration (e.g., \citealp{2014ApJ...783...91C,2015ApJ...798L..28C}). 
%In particular, \cite{2014ApJ...783...91C} have shown that quasi-perpendicular
%shocks are inefficient proton accelerators. However, in our case, it is difficult to imagine that 
%the magnetic field has a single orientation over Mpc scales in the upstream region of 
%cosmological shocks. Additionally, these simulations have been performed with several 
%simplifications because of computing limitations, and it remains to be seen if such results 
%will hold in 3D and for more realistic conditions.

%%%%%%%%%%%%%%%%%%%%%%%%%%%%%%%%%%%%%%%%%%%%%%%%%%%%%%%%%%%%%%%%%%%
\subsection{CR-to-thermal pressure}
Depending on the spectral and spatial distribution of the CR protons, the MAGIC upper limits from 
Table~\ref{tab:ULs} can be translated into constraints on the cluster CR population.
In Table~\ref{tab:pressure_UL}, we report the constraints on the CR-to-thermal 
pressure $X_\mathrm{CR}=P_\mathrm{CR}/P_\mathrm{th}$
obtained with the \emph{isobaric} model for different CR spectral indexes, 
$2.1\leq\alpha\leq2.5$, using the most constraining integral flux 
upper limit, i.e., the one above $1.6$\,TeV. 
The CR-to-thermal pressure must be below about 1\% for 
$\alpha\leq2.2$, below about 2\% for $\alpha=2.3$, and below 15\% for 
$\alpha=2.5$.
Including the EBL absorption effectively causes a worsening of the derived constraints of about $25$\%.
Therefore, the improvement of the constraints is modest, but the results are more robust with respect to \cite{2011arXiv1111.5544M}.

\begin{figure}[t!]
\centering
\includegraphics[width=0.5\textwidth]{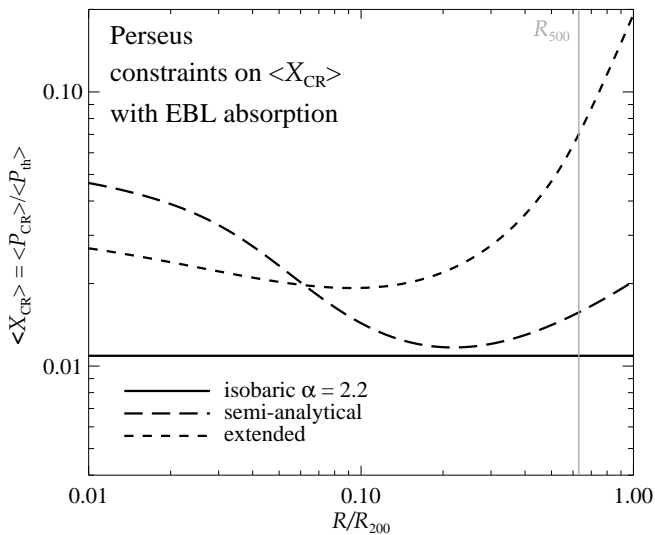}
\caption{Volume-averaged $X_\mathrm{CR}$ within a given radius $R$ as constrained by the upper limits 
presented in this work for the \emph{semi-analytical}, \emph{extended}, and \emph{isobaric} models, the last one 
with $\alpha=2.2$ matching the effective spectral index of the other models in the energies of interest here.
We recall that in the \emph{isobaric} model, $X_\mathrm{CR}(r)$ is constant by construction. The vertical grey
line represents the radius $R_{500}$, expressed with respect to an average density that is 500 times the critical density 
of the Universe, for an easier comparison with cosmological studies.}
\label{fig:XCR}
\end{figure}

The constraints on the volume-averaged CR-to-thermal pressure profiles  
for the \emph{semi-analytical} and \emph{extended} models are below 
1.4\% and 1.9\% within $0.15^\circ$ ($\approx 0.11 \times R_{200}$),
respectively, as shown in Figure~\ref{fig:XCR} and Table~\ref{tab:pressure_UL}. 
When considering the full galaxy cluster volume up to $R_{200}$, 
$\langle X_\mathrm{CR} \rangle$ is below 2\% for the \emph{semi-analytical} model, but 
significantly less constrained since below 19\% for the \emph{extended} model. This last
weak constraint is expected because the CR distribution is significantly flatter 
in the \emph{extended} model, leading to high $\langle X_\mathrm{CR} \rangle$ values in the 
cluster outskirts characterised by lower thermal pressure.

While in both simulations and analytical models, the CR distribution in galaxy clusters 
is assumed to roughly scale with the thermal gas, the real CR distribution is unknown 
and could be significantly flatter if CRs propagate out of the cluster core \citep{2011A&A...527A..99E,
2013arXiv1303.4746W,2012arXiv1207.6410Z}. Indeed, phenomenological evidence from observations of giant radio halos 
indicate that the CR distribution appears to be flatter than the ICM distribution independently of the generation mechanism
of the observed radio emission \citep{2012arXiv1207.3025B,2012arXiv1207.6410Z,2015arXiv150307870P}. 
This should be kept in mind when using galaxy clusters for cosmological purposes because, depending on the exact amount 
of CR protons, this can induce a bias in the estimation of cluster masses.
While 
this effect is limited to a few percent for the standard assumptions (\emph{isobaric} and \emph{semi-analytical}
models), a flat CR distribution could generate up to a 20\% bias (e.g., 
in the specific case considered here) in hydrostatic mass estimates, which are potentially relevant in the 
current era of precision cosmology.

\begin{table}[t]
\caption{Constraints on the volume-averaged CR-to-thermal pressure ratio within $R_{200}$ for different CR models.}
\begin{center}
\begin{tabular}{ccrrr}
\hline\hline
\phantom{\Big|}
model & $\alpha$ & $\langle X_\mathrm{CR,max}^\mathrm{no-EBL} \rangle$ [\%] & 
$\langle X_\mathrm{CR,max} \rangle$ [\%] \\
\hline\\[-0.5em]
\emph{isobaric} & $2.1$   &   0.5  &  0.7\\
                          & $2.2$   &   0.8  &  1.1\\ 
                          & $2.3$   &   1.7  &  2.3\\  
                          & $2.5$   & 11.4  & 15.2\\ 
\emph{semi-analytical} & $2.2$   &   1.5  &  2.0\\ 
\emph{extended}          & $2.2$   &   14.2  &  19.2\\                               
\hline
\end{tabular}
\end{center}
{\bf Notes.} Upper limits on the volume-averaged $X_\mathrm{CR}$ obtained within $R_{200}$
derived from the integral-flux upper limit in the energy range $1.6-10$\,TeV. 
$\langle X_\mathrm{CR,max}^\mathrm{no-EBL} \rangle$ is the constraint obtained neglecting the EBL absorption, given for comparison with previous results. 
$\langle X_\mathrm{CR,max} \rangle$ includes the EBL-absorption correction.
\label{tab:pressure_UL}
\end{table}

%%%%%%%%%%%%%%%%%%%%%%%%%%%%%%%%%%%%%%%%%%%%%%%%%%%%%%%%%%%%%%%%%%%
\subsection{Radio mini-halo and magnetic fields}
As mentioned in Section~\ref{sec:1}, the Perseus cluster hosts the brightest radio mini-halo 
known to date \citep{1990MNRAS.246..477P,Sijbring1993,2002A&A...386..456G}.
Assuming that the observed radio emission has a hadronic origin, i.e., is generated by 
secondary electrons produced in hadronic interactions between the CR and ICM protons,
the pion-decay $\gamma$-ray emission is directly linked to the radio signal.
Since the intensity of the synchrotron radio emission depends on the amount of secondary electrons, 
proportional to the hadronically-induced $\gamma$ rays, and the local magnetic field, our 
$\gamma$-ray upper limits can be turned into lower limits on the
cluster magnetic field \citep{2011arXiv1111.5544M}.

The $\gamma$-ray and synchrotron luminosities can be expressed as (adapted from \citealp{2008MNRAS.385.1242P})
\begin{eqnarray}
\label{eq:Lgamma}
L_\gamma &=& C_\gamma \dps\int d V\, n_\CR n_\rmn{ICM}\,,\\
   L_\nu &=& C_\nu \dps\int d V\, n_\CR n_\rmn{ICM} \frac{\dps
    \eps_B^{(\alpha_\nu+1)/2}}{\dps\eps_\rmn{CMB}+\eps_\rmn{SD}+\eps_B}\, ,
\label{eq:Lnu}
\end{eqnarray}
where $\alpha_{\nu}=\alpha/2$ is the synchrotron spectral index ($S_\nu \propto \nu^{-\alpha_\nu}$), 
$n_\rmn{CR}$ and $n_\rmn{ICM}$ are the CR and ICM densities, $C_\gamma$ and 
$C_\nu$ are constants that depend on the physics of the hadronic interactions,
and $\eps_B$, $\eps_\rmn{CMB}$ and $\eps_\rmn{SD}$ denote the energy density of the magnetic field (=$B^{2}/8\pi$), the CMB, and 
the star-and-dust light in the cluster, respectively. 
The magnetic field in galaxy clusters can be parametrised as
\begin{equation}
B(r) = B_0 \left( \frac{n_\mathrm{e}(r)}{n_\mathrm{e}(0)} \right)^{\alpha_{B}} 
\, ,
\end{equation}
where $B_0$ is the central magnetic field, $n_{e}$  the ICM electron density, 
and $\alpha_{B}$  a parameter describing the radial decline of the magnetic 
field. Such a parametrisation is favoured both by FR measurements in clusters and by 
hydrodynamical simulations (e.g., \citealp{2008A&A...482L..13D,2010A&A...513A..30B,2011A&A...529A..13K}).

Our radio synchrotron modelling includes energy losses due to IC scattering of ambient photons.
We consider the CMB, as well as the light from stars and dust (SD) in the cluster, $\eps_\rmn{SD}$,
according to the model of \cite{2011arXiv1105.3240P}, which has the advantage of including the average
contribution from the intra-cluster light obtained from a stacking of cluster observations.
We stress that the SD energy density dominates both the CMB and magnetic field
energy densities in the very centre of the cluster, typically $\leq 0.03\times R_{200}$,
for low values of the magnetic field, i.e., $B_{0} \lesssim 5$\,$\mu$G. Therefore, for the case of Perseus,
where the observed radio emission arises from within about $0.1\times R_{200}$ (see 
next section), including this term in the synchrotron losses can significantly
affect the modelling.

In Figure~\ref{fig:ConstrainedSpectra}, we show the EBL-corrected 
$\gamma$-ray emission within $0.15^\circ$ from the centre for both the 
\emph{semi-analytical} and the \emph{extended} models, corresponding to the parameters,
for which secondary electrons reproduce the observed radio surface brightness radial 
profile of the Perseus mini-halo at 1.4\,GHz 
\citep{1990MNRAS.246..477P}.\footnote{We take as reference here the radio surface 
brightness radial profile from \cite{1990MNRAS.246..477P} at 1.4\,GHz rather than 
the one at $327$\,MHz from \cite{2002A&A...386..456G}, as the latter may be 
affected by residual point-source contamination 
as pointed out in \cite{Sijbring1993} where the $327$\,MHz data was taken.}
We adopted three different values of the central magnetic field  $B_0 = 5$, $10$, and 
$20$\,$\mu$G with $\alpha_{B} = 0.3$ for the \emph{semi-analytical} model and 
$\alpha_{B} = 0.5$ for the \emph{extended} model as best-fit values.
Values of $\alpha_{B} \geq 0.5$ are theoretically preferred, for example
from simulations of gas sloshing in cool-core clusters \citep{2010ApJ...717..908Z,2011ApJ...743...16Z},
and the low value of $\alpha_B$ required for the \emph{semi-analytical} model 
is due to the centrally peaked CR profile in the \cite{2010MNRAS.409..449P} simulations,
in which AGN feedback was not accounted for, causing enhanced radiative cooling.

At parity of $B_0$, the \emph{extended} model always shows a higher $\gamma$-ray emission 
than the \emph{semi-analytical} model because of the flatter radial profile. In fact, 
while our $\gamma$-ray upper limits imply 
$B_{0} \gtrsim 5.5$\,$\mu$G for the \emph{semi-analytical} model, for the \emph{extended} model we obtain $B_{0} \gtrsim 8$\,$\mu$G
(for the choices of $\alpha_{B}$ as above). The FR measurements of cluster central magnetic field strengths range from 
$5$\,$\mu$G for merging clusters \citep{2010A&A...513A..30B,2013MNRAS.433.3208B}
to about $40$\,$\mu$G for cool-core objects \citep{2011A&A...529A..13K}. For the 
Perseus cluster, FR measurements are only available on very small scales, i.e., a 
few tens of pc, and this suggests magnetic field strengths of about $25$\,$\mu$G 
\citep{2006MNRAS.368.1500T}. There are, however, large uncertainties related 
to this measurement, and we refer the reader to more extensive discussions on this topic in our 
previous publications on the Perseus cluster 
\citep{2010ApJ...710..634A,2011arXiv1111.5544M}. 
In conclusion, when assuming the radio mini-halo in Perseus has a hadronic 
origin, the lower limits obtained on the magnetic field strength in the Perseus cluster 
using our $\gamma$-ray flux upper limits are consistent with FR measurements.

%%%%%%%%%%%%%%%%%%%%%%%%%%%%%%%%%%%%%%%%%%%%%%%%%%%%%%%%%%%%%%%%%%%
\subsection{Minimum $\gamma$-ray fluxes}
For clusters that host diffuse radio emission, such as the radio mini-halo in Perseus, 
we can estimate a theoretical minimum $\gamma$-ray flux in the hadronic scenario, 
which assumes that the observed radio emission has a secondary origin. The idea is that if the 
magnetic field is strong enough in all the radio-emitting region, i.e., 
$\epsilon_{B} \gg \epsilon_\mathrm{CMB} + \epsilon_\rmn{SD}$, 
a stationary distribution of CR electrons loses all its energy 
to synchrotron radiation \citep{2008MNRAS.385.1211P,2008MNRAS.385.1242P,
2010ApJ...710..634A,2011arXiv1111.5544M}. 
In this case, the ratio of $\gamma$-ray$-$to$-$synchrotron luminosity, $L_\gamma/L_\nu$,
becomes independent of the spatial distribution of CRs, of the ICM, 
and of the magnetic field, if the observed synchrotron spectral index is $\alpha_{\nu}=\alpha/2\approx1$,
as can be seen from Eqs.~(\ref{eq:Lgamma}) and (\ref{eq:Lnu}).
Therefore, a minimum theoretical $\gamma$-ray flux, $F_{\gamma,\,\mathrm{min}} = L_\gamma / 4 \pi D_\mathrm{lum}^2$, 
can be derived as
\begin{equation}
F_{\gamma,\,\mathrm{min}} \approx \frac{C_\gamma}{C_\nu} \frac{L_\nu}{4 \pi 
D_\mathrm{lum}^2} \, .
\label{eq:min_flux}
\end{equation}
A lower magnetic field value would require a higher secondary CR electron 
density, hence a higher CR proton density, to reproduce 
the observed radio synchrotron luminosity and would therefore result in a 
higher $\gamma$-ray flux. This is
why we can consider the above $\gamma$-ray flux as a theoretical 
minimum in the hadronic scenario.

The measured spectral index of the Perseus radio mini-halo ranges from 
$\alpha_\nu = 1.1$ to $\alpha_\nu = 1.4$  \citep{Sijbring1993,2002A&A...386..456G}. 
Assuming, for example, $B_{0} = 20$\,$\mu$G, the CR protons responsible 
for the GHz-synchrotron-emitting secondary electrons have an energy of $\sim$20\,GeV, 
which is about 400 times lower than the energy of the CR protons of 
$\sim$8\,TeV responsible for the TeV $\gamma$-ray emission. 
Here, again, we consider spectral indexes of $2.1\leq\alpha\leq2.5,$ which are 
consistent with the concavity in the CR spectrum that connects the low-energy CR population around GeV 
energies (characterised by $\alpha \simeq 2.5$) with the harder CR population at 
TeV energies ($\alpha \simeq 2.2$) (\citealp{2010MNRAS.409..449P}; see also 
the discussion in \citealp{2011arXiv1111.5544M}). 
While Eq.~(\ref{eq:min_flux}) is only approximately valid in the adopted range 
of spectral indexes, we estimate that the deviations in $L_{\gamma}/L_{\nu}$ from the exact equation 
with $\alpha = 2$ are within 10\%. 

We take as reference the total radio luminosity of the mini-halo measured at 
$327$\,MHz by \cite{Sijbring1993} of $L_{327\,\mathrm{MHz}} = 
17.57\times10^{-23}$\,erg\,s$^{-1}$\,Hz$^{-1}$\,cm$^{-2}$, and at $1.4$\,GHz by 
\cite{1990MNRAS.246..477P} of $L_{1.4\,\mathrm{GHz}} = 
3.04\times10^{-23}$\,erg\,s$^{-1}$\,Hz$^{-1}$\,cm$^{-2}$. 
The measured maximum emission radius of the Perseus radio mini-halo ranges from 
about 100\,kpc at 1.4\,GHz \citep{1990MNRAS.246..477P} to about 200\,kpc at 
327\,MHz \citep{Sijbring1993,2002A&A...386..456G}, which correspond to 
$0.075^\circ$ ($\sim$0.05$\times R_{200}$) and $0.15^\circ$ ($\sim$0.1$\times R_{200}$), respectively. 
The shape of the radio surface brightnesses at $327$\,MHz and at $1.4$\,GHz is more compact than the MAGIC PSF.
The minimum $\gamma$-ray flux of the hadronic scenario is therefore close to a point-like source for MAGIC.
About 67\% of the signal would be within the $\theta_{\rmn{cut}}$ used for the point-like assumption instead of 77\% (see Table~\ref{tab:Theta2Cuts}).
An appropriate correction factor of 1.15 must be applied to the point-like upper limits presented in Table~\ref{tab:ULs}
to correct for the expected emission not being perfectly point-like.
In doing so, we implicitly assumed that no $\gamma$-ray emission is coming from beyond the observed extension
in radio frequencies (see \citealp{2011arXiv1111.5544M} for a discussion).

\begin{figure}[t!]
\centering
\includegraphics[width=0.49\textwidth]{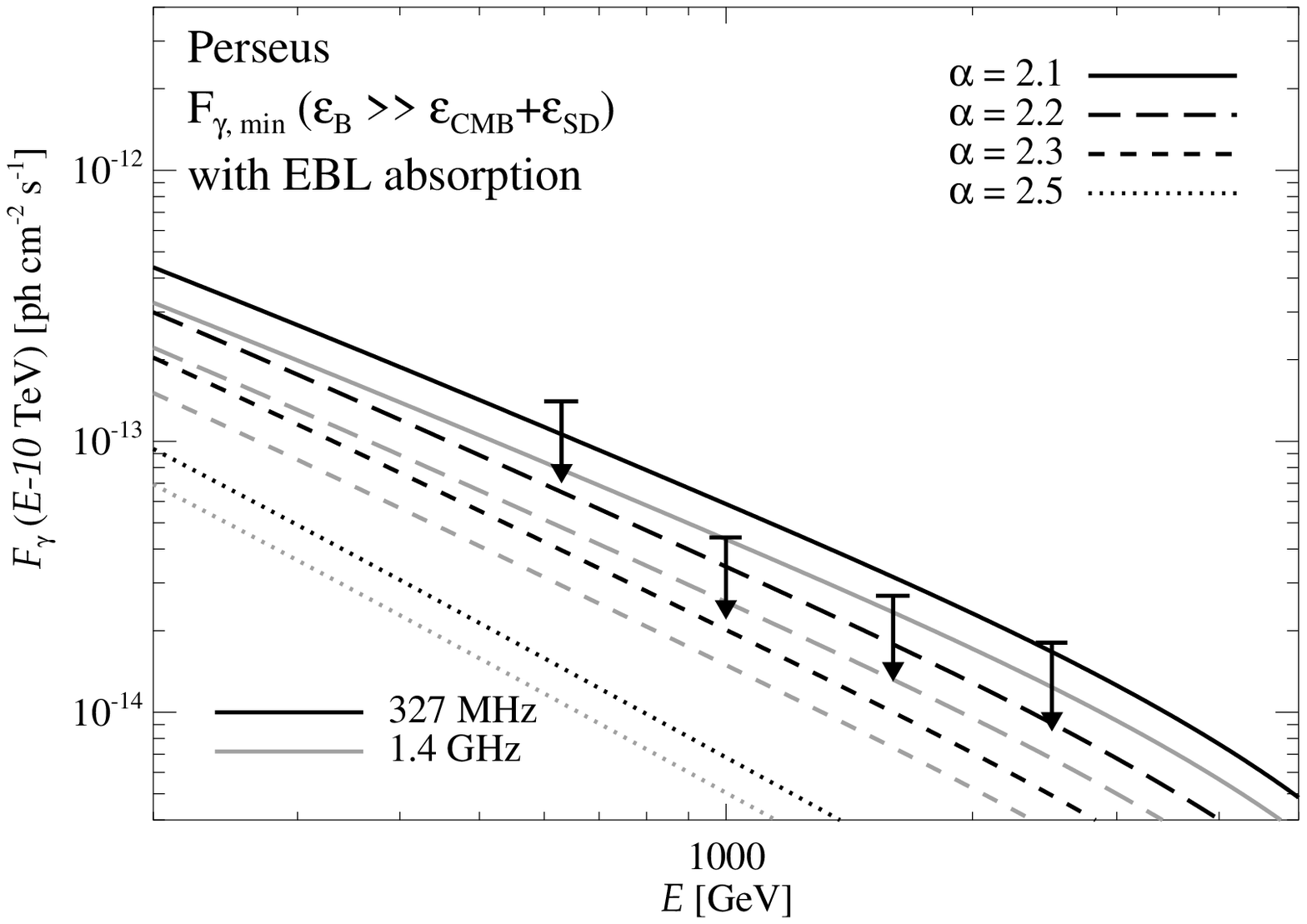}
\caption{Minimum $\gamma$-ray flux in the hadronic scenario obtained assuming 
that the Perseus radio mini-halo has a secondary origin
and that $\epsilon_{B} \gg \epsilon_\mathrm{CMB} + \epsilon_\mathrm{SD}$. 
We show the EBL-corrected fluxes for different 
spectral indexes $\alpha$, adopting as reference the total
synchrotron luminosity measured both at 327\,MHz \citep{2002A&A...386..456G} 
and at 1.4\,GHz \citep{1990MNRAS.246..477P}. 
We compare these with our 95\%-c.l. point-like upper limits appropriately
scaled for the radio surface brightness shape, i.e., multiplied by a factor of 1.15 with
respect to what is reported in Table~\ref{tab:ULs}. The flux
is integrated up to 10\,TeV.}
\label{fig:Fmin}
\end{figure}

Figure~\ref{fig:Fmin} shows the EBL-corrected minimum $\gamma$-ray emission 
derived as described above for $2.1\leq\alpha\leq2.5$, and adopting both $L_{327\,\mathrm{MHz}}$ and $L_{1.4\,\mathrm{GHz}}$ as synchrotron
luminosity. We compare
them with the properly scaled ($\times\,1.15$) point-like flux upper limits.
The most striking result is that the minimum $\gamma$-ray flux for $\alpha = 
2.1$ conflicts with our upper limits. This implies that, if $\alpha \le 2.1$, the observed 
diffuse radio emission in Perseus cannot be uniquely hadronic in origin, 
independently of the magnetic field strength in the cluster. 
For the softer spectral indexes considered here, the
current $\gamma$-ray upper limits cannot exclude the hadronic origin of the 
diffuse radio emission in Perseus. The other spectral index cases
should be in the reach of future ground-based $\gamma$-ray 
observations with the Cherenkov Telescope Array (CTA; e.g.,
\citealp{2013APh....43..189D}) with one order of magnitude better sensitivity.

%%%%%%%%%%%%%%%%%%%%%%%%%%%%%%%%%%%%%%%%%%%%%%%%%%%%%%%%%%%%%%%%%%%
%%%%%%%%%%%%%%%%%%%%%%%%%%%%%%%%%%%%%%%%%%%%%%%%%%%%%%%%%%%%%%%%%%%
\section{Conclusions}
\label{sec:6}
Clusters of galaxies are expected not only to contain $\gamma$-ray-bright AGNs, 
but also to host diffuse $\gamma$-ray emission due to neutral pion decays, induced by CR-ICM
hadronic interactions. Indeed, CR protons should be accelerated in 
clusters by structure formation shocks and injected by outflows from galaxies and AGNs,
and then should hadronically interact with the ICM protons, generating pions.
The most promising galaxy cluster to search for such diffuse $\gamma$-ray emission is the Perseus
cluster, which has been intensively observed with the MAGIC telescopes since 2008.
These observations resulted in the detection of two $\gamma$-ray-bright AGNs in the central galaxy NGC\,1275
and in the peculiar galaxy IC\,310, both already reported in previous MAGIC publications.
Here, we report the search of diffuse $\gamma$-ray emission, using 253\,hr of MAGIC observation in
stereoscopic mode, accumulated from 2009 to 2014.

The region of the hard point-like source IC\,310, located $0.6^\circ$ from the cluster centre,
can be easily excluded from our search for extended emission as this latter is expected to be centred on the
cluster core. The emission from the NGC\,1275 AGN, however, overlies the searched signal region.
We derived the most precise NGC\,1275 spectrum in the range 90\,GeV--1.2\,TeV ever done.
It is described well by a simple power law $f_{0}(E/200\,\mathrm{GeV})^{-\Gamma}$
with a very steep photon index $\Gamma=3.6\pm0.2_{\rmn{stat}}\pm0.2_{\rmn{syst}}$
and a differential-flux normalisation at 200\,GeV of
$f_{0} =(2.1 \pm 0.2_{\rmn{stat}} \pm 0.3_{\rmn{syst}}) \times 10^{-11} \mathrm{cm^{-2} s^{-1} TeV^{-1}}$,
in agreement with previous measurements.
No signal is detected above approximately 1\,TeV. Since the CR-induced emission is expected
to have a harder spectrum, we preferred the high energies for our search.
No other point-like emission is detected in the cluster, in particular from the radio galaxy NGC\,1265,
for which we derived a 95\%-c.l. flux upper limit above 250\,GeV of $5.6\times10^{-13}$\,cm$^{-2}$s$^{-1}$.

To bracket the uncertainty on the CR spatial and spectral distribution 
in Perseus, we considered three different models.
First, the \emph{isobaric} model, in which
the CR-to-thermal pressure is constant and the CR-spectrum index $\alpha$ ranges between 2.1 and 2.5.
Second, the \emph{semi-analytical} model of \cite{2010MNRAS.409..449P} 
was derived from hydrodynamical simulations of clusters, for which the CR spectrum follows a universal spectrum
with $\alpha=2.2$ at the energies of interest here.
Finally, the \emph{extended} hadronic model of \cite{2012arXiv1207.6410Z},
in which CRs propagate out of the cluster core and generate a significantly flatter radial distribution 
with respect to the previous two models. In this last model, the CR spectrum is the same as in the \emph{semi-analytical} model.
In this work we have not considered any softening of the CR-proton spectrum induced by possible CR propagation effects 
(e.g., \citealp{2013arXiv1303.4746W}).

We optimised our analysis for the different considered CR models.
No diffuse $\gamma$-ray emission or large-scale structures were detected in Perseus.
We derived 95\%-c.l. integral flux upper limits, in different energy ranges and compared to the 
signal expected from the models over the same range.
For the first time, we included the effect of the EBL absorption, which reduces the
$\gamma$-ray flux above 1\,TeV coming from Perseus by $\sim$20\%.
We discuss the fate of the produced electron pairs, including ICC and
possibility of plasma instabilities driven by the anisotropy of the pair beams.
We concluded that the absorbed $\gamma$ rays with $E>30$\,TeV are very likely reprocessed via ICC to our 
energy range of interest. In the most optimistic scenario (EGMF $< 10^{-14}$\,G), the ICC $\gamma$ rays could fully compensate for
the effect of the EBL absorption in 1--10\,TeV range.  
The strongest constraints on the CR models come from the 1.6--10\,TeV integral-flux 
upper limits of about $5 \times 10^{-14} \mathrm{cm^{-2} s^{-1}}$ 
in a central region of 0.15$^\circ$ radius.

The comparison with the \emph{semi-analytical} model sets a constraint on the maximum CR-proton acceleration efficiency, 
$\zeta_\mathrm{p, max}$, as defined in \cite{2010MNRAS.409..449P}.
The derived constraint, $\zeta_\mathrm{p, max}\le37$\%, is not much below our previous result obtained with 85\,hr of data
because the EBL absorption was not taken into account in that early work. Our new study is, therefore, more conservative
and more robust. We stress, however, that this constraint is only valid in the context of the
\cite{2010MNRAS.409..449P} model.

More model-independent constraints were set on the CR-to-thermal pressure ratio in the cluster.
In the context of the \emph{isobaric} model, $X_\mathrm{CR}$ must be $\la 1$\% for $\alpha\leq2.2$, 
$\la 2$\% for $\alpha=2.3$, and $\la 15$\% for $\alpha=2.5$. When considering 
the \emph{semi-analytical} model, $\langle X_\mathrm{CR} \rangle$ within $R_{200}$ is 
constrained to be $\la 2$\%. In the \emph{extended} model, $\langle X_\mathrm{CR} \rangle \la 2$\% within $0.15^\circ$,
but only $\la 20$\% within $R_{200}$ because the volume-averaged pressure ratio builds up
to significant values in the cluster outskirts where the ICM pressure drops. 
The actual CR distribution in clusters is unknown, and if it deviates significantly 
from the ICM distribution, as for example, in the \emph{extended} model, 
it could induce a bias on the estimates of the cluster hydrostatic mass, where its 
contribution is usually neglected, at a level that is potentially important in the current era 
of precision cosmology.

The Perseus cool-core cluster hosts the brightest known radio mini-halo.
Assuming that this diffuse radio emission is generated by synchrotron radiation
of secondary electrons from CR hadronic interactions with the ICM,
we can turn our $\gamma$-ray flux upper limits into lower limits
on the central magnetic field strength in the cluster.
For the first time, we included in our modelling energy losses due to 
IC scattering of ambient photons from stars and dust in the cluster, in addition to the commonly considered CMB.
We found $B_{0}\gtrsim 5.5$\,$\mu$G and $B_{0}\gtrsim 8$\,$\mu$G for the \emph{semi-analytical} and \emph{extended}
models, respectively. These constraints are consistent with FR measurements in clusters. 
Additionally, assuming that CR electrons lose all their energy by synchrotron emission in the radio emitting region
($\epsilon_{B} \gg \epsilon_\mathrm{CMB}+\epsilon_\rmn{SD}$),
the derived $\gamma$-ray flux becomes independent of the CR, magnetic field, and ICM distributions. 
This represents a theoretical flux lower limit in the hadronic scenario because lower magnetic fields would
imply a higher $\gamma$-ray emission.
With this approach, we found that for $\alpha\le2.1,$ the hadronic interpretation 
of the Perseus radio mini-halo is in conflict with our upper limits.
For more realistic $\alpha>2.1$, the minimum $\gamma$-ray flux is several times below our upper limits,
hence out of reach with MAGIC.  

The large amount of data presented in this work, about 250\,hr of observations, 
implies that it would be difficult to significantly improve upon our constraints with the current generation of Cherenkov telescopes.
Therefore, this five-year-long campaign
represents one of the reference results, together with the \emph{Fermi}-LAT 
observations, for the cluster physics in the $\gamma$-ray energy regime 
until the planned CTA observatory becomes operational in a few years from now.

%%%%%%%%%%%%%%%%%%%%%%%%%%%%%%%%%%%%%%%%%%%%%%%%%%%%%%%%%%%%%%%%%%%
%%%%%%%%%%%%%%%%%%%%%%%%%%%%%%%%%%%%%%%%%%%%%%%%%%%%%%%%%%%%%%%%%%%
\begin{acknowledgements}
The MAGIC collaboration would like to thank
the Instituto de Astrof\'{\i}sica de Canarias
for the excellent working conditions
at the Observatorio del Roque de los Muchachos in La Palma.
The financial support of the German BMBF and MPG,
the Italian INFN and INAF,
the Swiss National Fund SNF,
the ERDF under the Spanish MINECO (FPA2012-39502), and
the Japanese JSPS and MEXT
is gratefully acknowledged.
This work was also supported
by the Centro de Excelencia Severo Ochoa SEV-2012-0234, CPAN CSD2007-00042, and MultiDark
CSD2009-00064 projects of the Spanish Consolider-Ingenio 2010 programme,
by grant 268740 of the Academy of Finland,
by the Croatian Science Foundation (HrZZ) Project 09/176 and the University of Rijeka Project 13.12.1.3.02,
by the DFG Collaborative Research Centers SFB823/C4 and SFB876/C3,
and by the Polish MNiSzW grant 745/N-HESS-MAGIC/2010/0.
F.~Zandanel acknowledges the support of the Netherlands Organisation for 
Scientific Research (NWO) through a Veni grant.
\end{acknowledgements}

%%%%%%%%%%%%%%%%%%%%%%%%%%%%%%%%%%%%%%%%%%%%%%%%%%%%%%%%%%%%%%%%%%%
%%%%%%%%%%%%%%%%%%%%%%%%%%%%%%%%%%%%%%%%%%%%%%%%%%%%%%%%%%%%%%%%%%%
\bibliographystyle{aa}
\bibliography{bib_file}
%\bibliography{27846}

%%%%%%%%%%%%%%%%%%%%%%%%%%%%%%%%%%%%%%%%%%%%%%%%%%%%%%%%%%%%%%%%%%
%%%%%%%%%%%%%%%%%%%%%%%%%%%%%%%%%%%%%%%%%%%%%%%%%%%%%%%%%%%%%%%%%%
\label{lastpage}

\end{document}